\documentclass[reprint,amsmath,amssymb,braket,aps,longbibliography]{revtex4-2}

\usepackage{graphicx,color,braket,amsmath,dcolumn,bm}
\usepackage[colorlinks=true,linkcolor=blue]{hyperref}
\usepackage[T1,T2A]{fontenc}
\begin{document}

\title{A place for two-dimensional plasmonics in electromagnetic wave detection}

\author{Dmitry Mylnikov}%
\affiliation{%
 Laboratory of 2d Materials for Optoelectonics, Moscow Institute of Physics and Technology, Dolgoprudny 141700, Russia}%
\author{Dmitry Svintsov}%
 \email{svintcov.da@mipt.ru}
\affiliation{%
 Laboratory of 2d Materials for Optoelectonics, Moscow Institute of Physics and Technology, Dolgoprudny 141700, Russia}%

\begin{abstract}
Plasmons in two-dimensional electron systems (2DES) feature ultra-strong confinement and are expected to efficiently mediate the interactions between light and charge carriers. Despite these expectations, the electromagnetic detectors exploiting 2d plasmon resonance have been so far inferior to their non-resonant counterparts. Here, we theoretically analyse the origin of these failures, and suggest a proper niche for 2d plasmonics in electromagnetic wave detection. We find that a confined 2DES supporting plasmon resonance has an upper limit of absorption cross-section, which is identical to that of simple metallic dipole antenna. Small size of plasmonic resonators implies their weak dipole moments and impeded coupling to free-space radiation. Achieving the 'dipole limit' of absorption cross-section in isolated 2DES is possible either at unrealistically long carrier momentum relaxation times, or at resonant frequencies below units of terahertz. We further show that amendment of even small metal contacts to 2DES promotes the coupling and reduces the fundamental mode frequency. The contacted resonators can still have deep-subwavelength size. They can be merged into compact arrays of detectors, where signals from elements tuned to different frequencies are summed up. Such arrays may find applications in multi-channel wireless communications, hyper-spectral imaging, and energy harvesting.  

\end{abstract}

\pacs{Valid PACS appear here}
\maketitle

\section{Introduction}
Two-dimensional plasmons represent a remarkable example of electromagnetic waves confined to scales two orders of magnitude below the free-space photon wavelength~\cite{Stern1967}. It is commonly articulated that confinement allows 2d plasmons to mediate strong light-matter interactions~\cite{Koppens2011,Grigorenko2012,Abajo_GraphenePlasmonics}. What is meant by this mediation remains largely unclear. The fundamental studies reveal intriguing properties of 2d plasmons~\cite{Ni_LimitsToPlasmonics_2018,Woessner2015,Faist_UltraStrongCoupling,Alonso-Gonzalez2014}, yet do not show a distinct promise for optoelectronics.

This long-reigning existential vacuum was filled by a proposal to use 2d plasmons for high-responsivity electromagnetic detection~\cite{Dyakonov1996}. The idea partly followed the path of bulk plasmonics, with its established applications in photodetectors and solar cells~\cite{Atwater2010}. More specifically, Ref.~[\onlinecite{Dyakonov1996}] suggested a confined 2d electron system (2DES) to act simultaneously as a resonant cavity and rectifier (detector) for 2d plasmons. The estimated detector response to free-space illumination scaled quadratically with plasmon quality factor, and promised sensitive detection for practically important terahertz range~\cite{MIttleman_NatEl_THz}. The idea ignited a booming research in the field of 2d plasmon-enhanced detectors, with further proposals based on different materials~\cite{Tredicucci_DeviceConceptsGraphene,Viti_Detection_TI_Surface_states,Viti_BlackP,Yavorsky_HgTe_detection,Tomadin_PRB_PlasmawaveResponse,Principi_PRB_CorbinoDetector}, rectification mechanisms~\cite{Ryzhii_Shur_JJAP,Popov_rectification_plasma_waves,Pseudo_Euler,Kachorovskii_performance_limits}, and coupling schemes~\cite{Olbrich2016,Meziani_ADGG_graphene}.

So far, the {\it experimentally realised} 2d plasmonic detectors~\cite{Dorozhkin2005,Muravev2012a,Knap_resonant, Peralta2002, Cai_photocurrent_by_plasmons,Freitag_Intrinsic_plasmons,muravev2016_interfermometer,Chudow_Plasmons_CNTs,Bandurin_resonant} did not demonstrate substantial responsivity enhancement compared to their non-resonant counterparts~\cite{Castilla2019,Viti_hBN_encapsulated_Low_noise,Gayduchenko2021,Bauer2019}. Even if plasmonics was in play, and responsivity demonstrated clear maxima at plasmon frequencies~\cite{Dorozhkin2005,Muravev2012a,muravev2016_interfermometer,Bandurin_resonant}, their height was not impressive for practical applications. Electronic quality of 2d systems and extra mechanisms of plasmon loss were typically blamed as limiting factors to the strength of plasmon resonance~\cite{Knap_resonant}. But even in cleanest 2d systems with electron mobility exceeding millions centimeters squared per volt per second, excitation of plasmons lead not to resonant enhancement of responsivity, but to responsivity oscillations~\cite{Muravev2012a,muravev2016_interfermometer}.

These numerous experiments stimulate a critical re-consideration of advantages which 2d plasmonics can bring to electromagnetic wave detection. In this paper, we'll show that plasmonic detectors based on confined 2DES possess a fundamental limit of electromagnetic absorption cross-section order of $\sigma_\lambda = \lambda_0^2/4\pi$, where $\lambda_0$ is the free-space wavelength~\cite{tretyakov2014maximizing}. This limit is identical for resonant plasmonic detectors and non-resonant structures matched to conventional antennas~\cite{balanis2015antenna}; hence, a single plasmonic detector can't help increasing responsivity. Further, reaching this maximum cross-section in plasmonic systems is far more challenging, compared to antenna-coupled detectors. The reason is small size of 2D plasmonic resonators, their small dipole moment, and weak coupling to free-space radiation.

\begin{figure*}[ht] \centering
    \includegraphics[width=\textwidth]{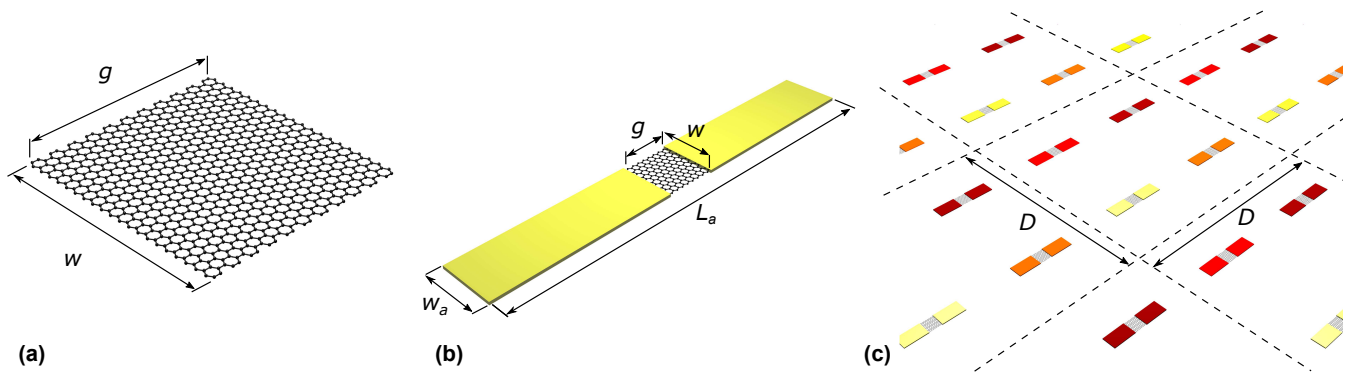}
    \caption{{\bf Two-dimensional plasmonic resonators} (a) Isolated layer of two-dimensional electronic system (graphene is shown as example) (b) 2DES connected to metallic leads (c) Multi-colour array of plasmonic detectors}
    \label{fig1}
\end{figure*}

We explore the ways to achieve perfect coupling between confined 2d plasmons and free-space radiation. We find that metal contacts, being an inevitable part of any photodetector, strongly modify the absorption cross-section, matching conditions, and resonant frequencies. At fixed size of 2DES, the electron mobility providing best matching scales inversely proportionally to the square of contacts' length. 
Loosely speaking, a good plasmonic detector should have carefully engineered metal contacts, but not the highest-mobility 2d channel. 

Despite the emerging identity between responsvity of plasmonic and non-plasmonic detectors, we find a suitable niche for 2d plasmonic detectors. Exploiting the deep-subwavelength confinement, one can 'pack' a large number of detectors even in the minimally-focused light spot size of area $\sim \lambda_0^2$. Each detector can have absorption cross-section reaching $\sigma_\lambda$. Thus, the net signal collected from an array can largely exceed the signal from one non-plasmonic antenna-matched device (having a comparable area). Importantly, additive character of cross-section works if only plasmonic detectors within a spot are tuned to slightly different resonant frequencies. Applications of such detectors can lie in hyper-spectral imaging~\cite{Kivshar_Hyperspectral},  multi-channel wireless communication~\cite{Nagatsuma2016}, and electromagnetic energy harvesting~\cite{Sharma_rectenna}. 

So far, the impedance matching requirements for confined 2d plasmons haven't yet been realised, though they received much attention in metal-insulator and nanoparticle plasmonics~\cite{Spinelli_ImpedanceMatching_2011,Ginzburg_impedance_matching_2007}. Recently, it was shown that short wavelength of {\it propagating} 2d plasmons implies their low excitation efficiency by small scatterers~\cite{Abajo_Limits_to_Coupling}. Interplay of Ohmic and radiative damping was recently studied for 2d plasmons in discs~\cite{Zagorodnev-Effect-of-retardation}, but practically important absorption cross section were not analysed. General constraints on extinction and absorption were recently derived in [\onlinecite{Kuang_MaxOpticalResponse}], but applications were demonstrated only for bulk plasmonic structures. The absorption cross sections limited by material loss in 2DES were analysed in [\onlinecite{Miller_MaxOpticalResponse2dm}], but developed quasistatic theory did not reproduce the dipole cross-section limit.

Further on, we shall theoretically elucidate the strategies for making a plasmonic photodetector with maximum radiation absorption cross-section. We do not focus on particular photocurrent generation mechanism (be it thermoelectric, bolometric, photovoltaic, or resistive self-mixing), as all they are bounded by the absorbed power. The simulations are performed with CST Microwave Studio package, and supported by simple analytical estimates if possible.

\section{Absorption limits by confined 2d plasmons}
We start by examining the electromagnetic absorption by isolated two-dimensional systems without metal contacts [Fig.~\ref{fig1}(a)]. This case, being not much practical for detectors, provides useful insights into matching conditions. The properties of 2DES are described in terms of kinetic inductance per square $L_\square$ and active resistance $R_\square$, such that sheet impedance $Z_\square = i \omega L_\square + R_\square$. The frequency dependence of $Z$ can be arbitrary, and may originate from intra- and interband material processes; all we need in simulations is just numerical value of $Z$ at excitation frequency $\omega$. If the conductivity is described by Drude model, $L_\square$ is linked to sheet carrier density $n_s$ as $L_\square = m/n_s e^2$, and resistance $R_\square = m/n_s e^2\tau$, where $\tau$ is the momentum relaxation time. Ths 2DES has a square shape with side $g$.

\begin{figure*}[ht] \centering
    \includegraphics[width=\textwidth]{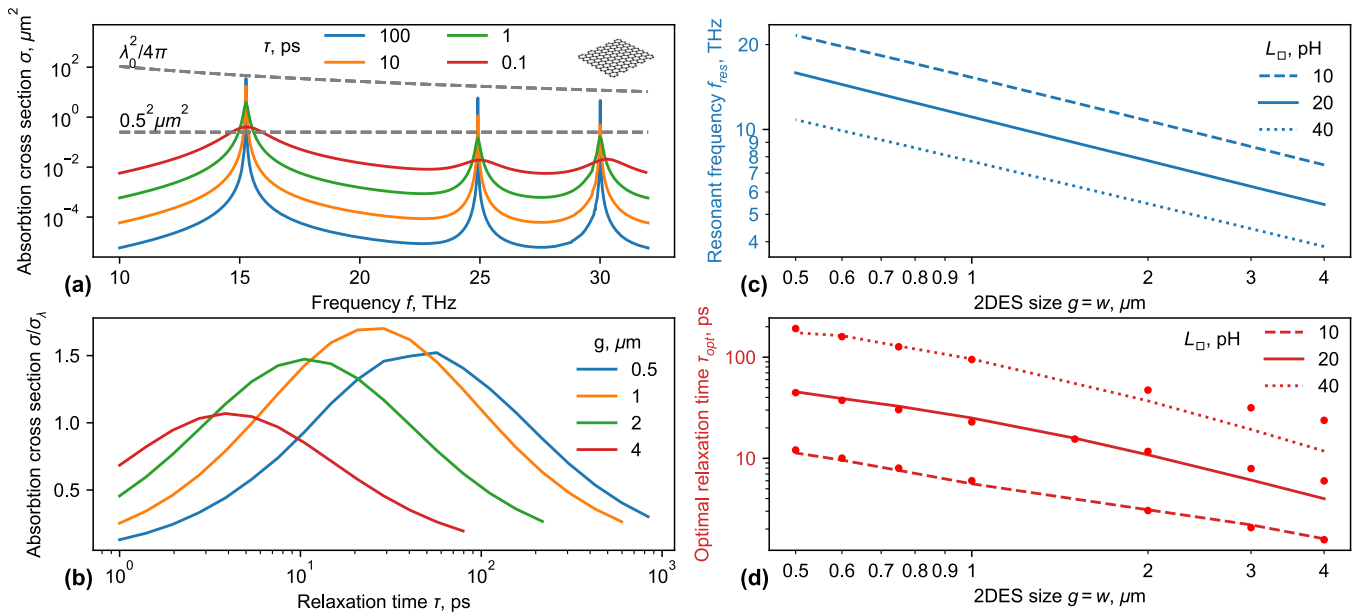}
    \caption{{\bf Electromagnetic absorption by confined 2d plasmons} (a) Absorption cross-section section spectrum of a square 2DES sheet ($g=w=0.5$ $\mu$m) at different momentum relaxation times $\tau$. Sheet inductance $L_\square=20$ pH (b) Absorption cross-section at fundamental plasmonic resonance vs momentum relaxation time at variable size $g$: emergence of optimum momentum relaxation $\tau_{\rm opt}$ (c,d) Dependence of resonant frequency $f_{\rm res}$ and matching time $\tau_{\rm opt}$ on 2DES size $g$ for different values of sheet inductance $L_\square=10$, 20 and 40 pH (controlled by carrier density). Dots in (d) are the results of numerical simulations, lines are calculated using analytical expression (\ref{eq:match_isolated})
     }
    \label{fig:Isolated_graphene}
\end{figure*}

The calculated absorption spectrum of a square piece of 2DES ($g=0.5$ $\mu$m) with kinetic inductance $L_\square = 20$ pH and moderate momentum relaxation time $\tau = 0.1$ ps is shown in Fig.~\ref{fig:Isolated_graphene}. The parameters correspond to graphene with Fermi energy $E_F \approx 0.4$ eV and mobility $\mu = 2.4 \times 10^3$ cm$^2$/V s~\footnote{For GaAs quantum well with $m^*=0.067m_0$ this inductance corresponds to sheet density $n_s = 1.2 \times 10^{13}$ cm$^{-2}$, while $\tau = 0.1$ ps corresponds to mobility $\mu \approx 2600$ cm$^2$/V s.}. The plasmon resonance at $f_{\rm res} \approx 15$ THz manifests as a clear absorption maximum, as the high-quality conditions are well fulfilled, $\omega_{\rm res} \tau = 10 \gg 1$. The cross section largely exceeds the geometrical area of 2DES $\sigma \approx 10 g^2$.

Would one obtain larger resonant absorption if the electronic quality of 2DES would be increased? We examine it by evaluating the cross-sections for a sequence of relaxation times, $\tau = 1$, $10$ and $100$ ps. The resonant absorption $\sigma(f_{\rm res})$ initially grows, reaches a maximum at $\tau_{\rm opt} \approx 50$ ps, and then rolls down. The fundamental limits of cross-section are best visualised if plotted in units of $\sigma_\lambda = \lambda_0^2/4\pi$ [Fig.~\ref{fig:Isolated_graphene} (b)]. The cross-section at optimum relaxation time is order of $(1...1.5) \sigma_\lambda$, which is a characteristic of dipole resonance limited by radiative decay~\cite{tretyakov2014maximizing}.

The optimum relaxation time $\tau_{\rm opt} \sim 10... 100$ ps for reaching the best absorption at terahertz frequencies is quite large for realistic 2DES, as it corresponds to mobilities $\mu_{\rm opt} \approx (0.2 ...2)\times 10^6$ cm$^2$/(V s). The very existence of optimum time hints that the resonant absorption maximum corresponds to some kind of matching. Further on, the optimum relaxation time becomes longer once the size of 2DES is decreased and its resonant frequency goes up. It suggests that matching problem is associated with small size of plasmonic resonator, short wavelength of 2d plasmons, and ultimately weak radiation coupling of plasmonic systems.

To test this idea and get a simple estimate of optimum relaxation time, we equate the radiation losses $P_{\rm rad}$ and Joule losses $P_{\rm J}$ in a piece of 2DES with high-frequency current distribution ${\bf j}_\omega({\bf r})$ (Appendix \ref{app:matching}). This results in
\begin{equation}
\label{eq:match_isolated}
    \frac{4\pi}{3}\left(\frac{g}{\lambda_0}\right)^2 Z_0 \langle {\bf j}_\omega({\bf r}) \rangle^2 = 2R_\square \langle {\bf j}^2_\omega({\bf r}) \rangle,
\end{equation}
where $Z_0 = 377$ $\Omega$ is the free-space impedance, and angular brackets $\langle ... \rangle$ denote averaging over the sample area. For a fundamental mode with nearly-sinusoidal current distribution one can estimate $\langle {\bf j}^2_\omega({\bf r}) \rangle / \langle {\bf j}_\omega({\bf r}) \rangle^2 = \pi^2/8 \sim 1$. The inverse quadratic scaling of radiated power with wavelength is in agreement with recent experimental measurements~\cite{Muravev_FineStructureCR}.

The matching equation (\ref{eq:match_isolated}) has serious implications. The value of $\tau_{\rm opt}$ should be so large that Ohmic resistance $R_\square$ is $(\lambda_0 / g)^2$ times smaller than free-space impedance. We note that sheet resistances below $Z_0$ were reached only recently in high-quality GaAs quantum wells~\cite{Muravev_PRL_Relativistic}, while typical graphene sheets have $R_\square \sim 1$ kOhm. The factor $(g/\lambda_0)^2$ requiring the smallness of Ohmic resistance is nothing but the measure of sub-wavelength compression achieved in a plasmonic resonator.

Scaling of optimum relaxation time with size of plasmon resonator $g$ can be derived once the scaling of resonant frequency is known. As an estimate, one can ''quantize'' the dispersion of 2d plasmons in extended system $\omega_{\rm pl}(q) = \sqrt{q/2\varepsilon_0 L_\square}$ at the characteristic wave vector $q = \pi/ g$. As a result, $f_{\rm res} = (2\pi)^{-1} \omega_{\rm pl}(\pi/g)$ becomes inversely proportional to the square root of size, $f_{\rm res} \propto g^{-1/2}$, while optimum momentum relaxation time scales approximately as $1/g$. These trends are supported well by our numerical simulations demonstrated in Fig.~\ref{fig:Isolated_graphene} c and d.

The inverse scaling of matching time $\tau_{\rm opt}$ with size of plasmonic structure $g$ suggests that strong light absorption and high-responsivity detection may be achieved only at THz and sub-THz frequencies with very high quality of 2DES. Already for mid-infrared applications ($\lambda_0 \approx 3$ $\mu$m), the matching time $\tau_{\rm opt} \sim 2$ ns exceeds the achievable values by three orders of magnitude. We shall further show that these severe matching conditions can be modified via engineering of metal contacts to 2DES. 

\section{Matching 2d plasmons to free space with metal contacts}

\begin{figure*}[ht!] \centering
    \includegraphics[width=\textwidth]{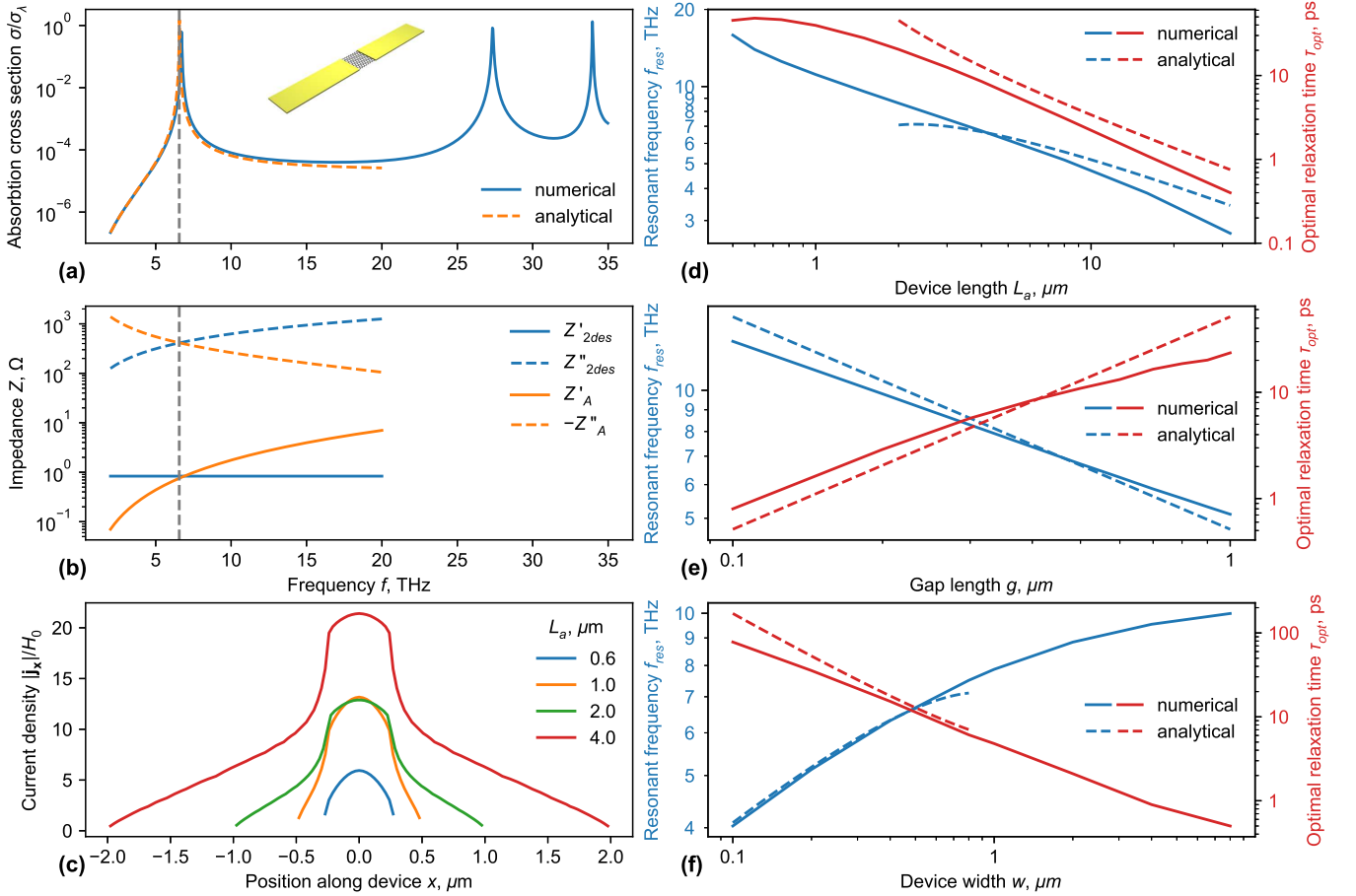}
    \caption{{\bf Engineering the metal leads for coupling 2d plasmons to free-space radiation}
    (a) Absorption cross section of 2DES sheet of size $g=w = 0.5$ $\mu$m with metal arms. Total device length $L_a=4$ $\mu$m, inductance $L_{\square}=20$ pH, $\tau=\tau_{opt}\approx12$ ps. (b) Frequency scaling of 2DES (blue) and antenna (orange) impedance in the vicinity of fundamental resonance (c) Distribution of induced currents $j_x(x,y=0)$ along the contacted 2DES at resonance with increasing the length of metal arms. $\tau=1$ ps (d-f) Scaling of resonant frequency and optimum momentum relaxation time in contacted 2DES as functions of device length $L_a$ (d), 2DES size $g$ (e) and net device width $w$ (f) }
    \label{fig3}
\end{figure*}

Metal contacts are inevitably present in any semiconductor-based photodetector, and may affect the device electrodynamics. Previously, a strong effect of contacts on eigenmodes of 2d plasmons was demonstrated experimentally~\cite{Muravev_PhysicalOrigin,Muravev_LCmode}. Here, we show that metal contacts of moderate length have a strong impact on radiative coupling of plasmonic devices, and considerably reduce $\tau_{\rm opt}$.


Figure \ref{fig3} substantiates our suggestions by showing the spectrum of cross section (a), the resonant frequency, and optimum momentum relaxation time (c) for 2DES with finite-length metal arms. The first example shown in Fig. \ref{fig3}(a) shows that amendment of metal arms to square-shaped 2DES with moderate relaxation time $\tau = 12$ ps considerably enhances the cross-section. Eventually, it can become close to the fundamental limit of $\sigma_\lambda$.

It is instructive to track the properties of fundamental plasmonic mode of 2DES with increase in metal arm length $L_a$. Our simulations show that resonant frequency scales approximately as $f_{\rm res} \propto L_a^{-1/2}$, while 'matched' relaxation time scale as $\tau_{\rm opt} \propto L_a^{-1}$ if the arm length exceeds the size of 2DES, $L_a \gg g$. Already for $L_a = 10$ $\mu$m and $g \approx 0.5$ $\mu$m, the optimum momentum relaxation time can take quite a realistic value $\sim 3$ ps. For a similar contact-less 2DES, the matched $\tau_{\rm opt}$ is roughly an order of magnitude longer.

The above comparison of matching conditions in contacted and contact-less 2DES was performed at given 2DES size. Of course, larger-area 2DES also have shorter relaxation time, just as small 2DES with large metal contacts. Comparing the values of $\tau_{\rm opt}$ in contacted and contact-less 2DES {\it at fixed resonant frequency}, we find them to be nearly identical. From viewpoint of practical photo-detection, it is desirable to use small 2DES with large metal arms, as this leads to enhancement of power density released in 2d channel.

It is further possible to provide a simple model governing the matching of plasmonic resonators with the aid of contacts. A hint toward such a model lies in a very simple distribution of currents in contacted 2DES. While uncontacted 2DES at resonance has a nearly sinusoidal current distribution, this profile is quickly flattened toward constant value as $L_a$ goes up [Fig.~\ref{fig3} (c)]. It implies that effects of distributed inductance and capacitance are unimportant for contacted resonators, and simple lumped models can work well.

The desired lumped model is simply a series connection of 2DES with impedance $Z_{\rm 2des} = (g/w)Z_\square$ and short metal dipole antenna with with impedance $Z_A  = R_{\rm rad} + (i \omega C_A)^{-1}$. Imaginary part of 2DES impedance is mainly due to kinetic inductance of 2D electrons. The antenna impedance is mainly governed by its capacitive gap. The resonant frequency in this case is simply given by~\cite{Muravev_LCmode}
\begin{equation}
    \omega_{0} = \frac{1}{\sqrt{L_{\square} C_a}} \sqrt{\frac{w}{g}}.
\end{equation}
To complete the model, we adopt the expressions for capacitance and radiative resistance of a short dipole antenna~\cite{balanis2015antenna}, $C_a \approx \frac{\pi}{2}\varepsilon_0 L_a \ln^{-1}[L_a/w_a-1] $ and $R_{\rm rad} =\frac{\pi}{6} Z_0 (L_a/\lambda_0)^2$. The matching condition, in such model, reads simply as $R_\square g/w = R_{\rm rad}$, or, in extended form:
\begin{equation}
    R_\square \frac{g}{w} = \frac{\pi}{6} Z_0 \left(\frac{L_a}{\lambda_0}\right)^2.
\end{equation}

The above analytical expressions confirm the numerically observed scaling of resonant frequency and matching time, and correct them by a logarithmic factor $f_{\rm res} \propto [L_a/ \ln(L_a/w_a)]^{-1/2}$ and $\tau_{\rm opt} \propto [L_a \ln(L_a/w_a)]^{-1}$. The lumped theory predicts the values of $f_{\rm res}$ and $\tau_{\rm opt}$ very well, as seen from Fig.~\ref{fig3} (d), (e) by comparison of solid and dashed curves. Another option for increased matching of 2d plasmonic resonators lies in increase of arms' width $w$. As soon as $w \ll \lambda_0$, increased width leads to quadratic enhancement of radiating dipole moment. This makes $\tau_{\rm opt}$ drop abruptly as $w$ increases, as shown in Fig.~\ref{fig3} (f). Ammending the antenna reciprocity theorem to the lumped model (Appendix \ref{app:reciprocity}), one can further analytically predict the value of $\sigma$, as demonstrated in Fig.~\ref{fig3} with dashed line.

The above matching options enable, in principle, achieving large absorption cross-sections by plasmonic photodetectors while keeping their deep sub-wavelength confinement. Indeed, the size of optimally-matched contacted structures in considered in Fig.~\ref{fig3} did not exceed $0.5$ $\mu$m $\times 4$ $\mu$m. Small areas of plasmonic detectors allow sparing the space on a chip, compared to large antenna-coupled detectors. This is a purely economic (yet important) benefit. Still, the absorption cross-sections (and responsivities) of a single plasmonic and a single antenna-coupled non-plasmonic detectors are bounded by the same values.  

\begin{figure}[h] \centering
    \includegraphics[width=0.5\textwidth]{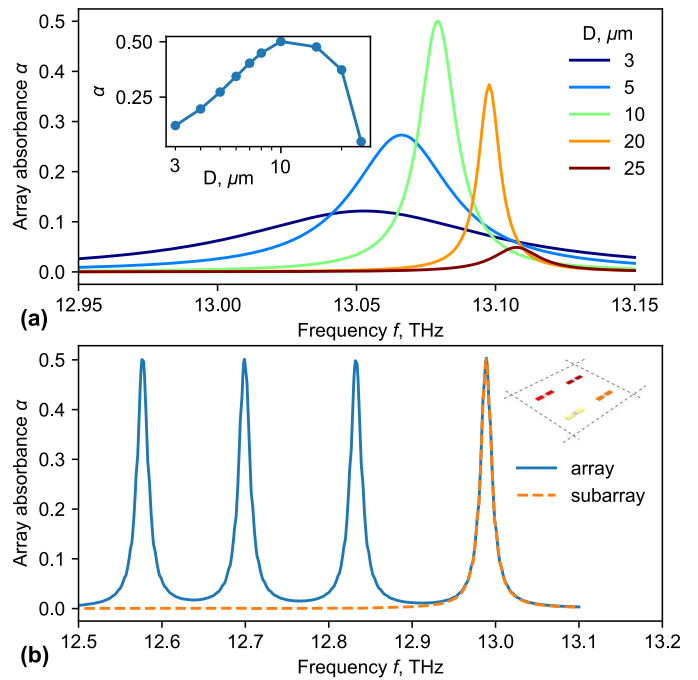}
    \caption{{\bf Single and multi-color arrays of plasmonic photodetectors} (a) Absorbance spectrum $\alpha(f)$ of a single-colour array, i.e. a square-lattice of of identical plasmonic detectors. Different lines correspond to different cell periods $D$. Inset: resonant absorbance vs period (b) Absorption spectrum of the four-colour array with optimum cell period $D=10$ $\mu$m. The absorbances of individual detectors are non-interfering and each reaches maximum value of $0.5$. Inductance $L_{\square}=20$ pH, 2DES lengths $g=300,\,310,\,320,\,330$ nm, width $w=500$ nm, total length $L_a=1$ $\mu$m, $\tau=\tau_{opt}=23$ ps}
    \label{fig4}
\end{figure}

\section{Exploiting deep-subwavelength confinement for photodetection}

While a single 2d plasmonic resonator and an optimally-loaded dipole antenna have the same absorption cross-sections, it may be tempting to exploit the small size of plasmonic cavities for design of detector arrays. By connecting the devices in series (or in parallel), one may progressively increase the photovoltage (or photocurrent) generated by such an assembly. One can estimate the maximum number of plasmonic devices per optimally focused light spot as $N \sim (\lambda_0/\lambda_{\rm pl})^2$. Thus, one may expect an $N$-fold enhancement of plasmonic meta-device responsivity, compared to non-plasmonic device using classical antenna focusing. 

Such packing is hardly possible if all detectors are tuned to the same resonant frequency. The emerging complexities for plasmonic arrays are illustrated in Fig.~\ref{fig4}. Here, we plot the absorbance $\alpha(f)$ by a square-lattice array of plasmonic detectors tuned to $f_0\sim 13$ THz vs the lattice period $D$. The absorbance is defined as the cross-section per unit cell area, $\alpha = \sigma/D^2$. As the packaging becomes denser, the absorbance first goes up ($D \lesssim 10$ $\mu$m). By further increasing the density, we observe a strong degradation of $\alpha$.  Instructively, the optimal cell period appears to be order of $\sigma^{1/2}_\lambda \approx 6.5$ $\mu$m. It implies that plasmonic device at a resonance strongly distorts the field in a large nearby area $\sim \sigma_\lambda$, and exploiting the 'compactness' advantage of plasmonics becomes impossible for single-frequency detection. Thus, the 'effectively occupied area' by plasmonic cavity is still order of $\sigma_\lambda$, and hence order of size of a classical antenna. 

The situation becomes more encouraging once we recall that large absorption cross-section occurs only at the resonant frequency. One may thus use the $\sigma_\lambda$-circle near the 2d plasmonic resonator for placing other detectors tuned to different frequencies. Dissimilar devices do not hinder each other's absorption, provided their line width $\delta \omega \sim \tau^{-1}_{\rm opt}$ exceeds the spacing between resonant frequencies. Figure~\ref{fig4} (b) substantiates this idea and compares the absorbance of single-color array (orange line) with that of four-color array of the same period (blue line). Though formal lattice periods for these arrays are equal, the area per element in multi-colour case is four times smaller. Each of four tuned devices fully exploits its resonant cross-section.

We may suggest three application domains for such detector arrays. First of all, they can be exploited for sensitive detection of signals comprised of several wavelengths. This situation is common for multi-channel wireless communications. The plasmonic detector array resonantly responding to all $M$ communication wavelengths occupies the space $\lesssim \sigma_\lambda$. Achieving the same result with classical dipole antennas would require the area of $M \sigma_\lambda$, which becomes highly impractical for $M \gg 1$. Second, the arrays of plasmonic detectors connected in series can effectively generate photovoltage upon illumination by broadband signals. The limiting example of such broadband signal is the background radiation noise, nowadays considered as a source for energy harvesting. Finally, plasmonic arrays can find applications for compact hyper-spectral imagers, where signal from each resonating element is proportional to light intensity at given frequency. 

\section{Discussion and conclusions}

The presented study was aimed to indicate a possible niche for 2D plasmonics in photodetection applications. We tried to dispel a common view that high quality factor for 2d plasmons is a necessary and sufficient condition for making a good plasmonic photodetector. Instead, we have shown that a good plasmonic photodetector requires impedance matching between radiative and Ohmic resistances. This is most simply achieved by engineering metal contacts to 2DES. In some situations (like at the mid-infrared frequencies), large electron momentum relaxation time are indeed essential. The quality requirements to 2DES are, however, becoming less important for THz and sub-THz frequencies.

In our study, we have omitted the consideration of photodetectors based on grating-coupled 2DES~\cite{Peralta2002,Olbrich2016}. For an infinite grating, the maximum absorbance at plasmon resonance does not exceed 50 \%, which is the same as for 2d film perfectly matched to free space. The gain from plasmonic effects here is not as essential as for confined 2DES. Absorption cross-sections by finite grating-coupled 2DES~\cite{Meziani_ADGG_graphene} requires a separate consideration, but we expect the matching requirements similar to Eq.~(\ref{eq:match_isolated}) to appear in that case. For the same reason, we haven't considered the 2d photodetectors with arrays of plasmonic particles for enhanced absorption~\cite{Echtermeyer2011,Shalaev_2020}. The latter also have no relation to 2d plasmonics.

It would be supercilious to say that our proposed multi-color arrays based on 2D plasmons are an only niche for 2D plasmonics in photodetection. Of course, future inventions can't be predicted. Ultra-high confinement is not the only important property of 2d plasmons, another to be mentioned is the electric and magnetic tunability of resonant frequency. The latter enables the design of all-electrical radiation spectrometers. High responsivity and tunability are essential for spectroscopy, while deep-subwavelength size is not essential. In this niche, it may be beneficial to use 2DES with moderate mobility optimally matched with large ($\sim \lambda_0$) antenna arms. It is possible to show (Appendix \ref{app:spectrometer}) that tunability of resonant frequency persists in this case, while relaxation time requirements are softened.

It is tempting to reveal why previous proposals of the resonant plasmonic detectors~\cite{Dyakonov1996,Tomadin_PRB_PlasmawaveResponse,Ryzhii_Shur_JJAP,Principi_PRB_CorbinoDetector} strongly overestimated their responsivities. The reason lies in assumption that incident electromagnetic field can be mapped to a perfect voltage source with zero internal resistance. Assuming very long momentum relaxation time, one could obtain arbitrarily large $Q$-factor of plasmon resonance. In reality, radiative resistance $R_{\rm rad}$ of all plasmonic systems is finite. Moreover, radiative resistance is linked to the ability of detector to receive power from free space. By releasing the assumption of voltage source ideality, one immediately realizes the existence of limiting responsivity  for plasmonic radiation detectors.

This work was supported by the grant MK-1035.2021.4 of the President of Russian Federation. The authors thank V. Muravev for fruitful discussions.

\appendix
\section{Matching the absorbed and radiated power}
\label{app:matching}
We derive the matching condition by equating the absorbed ($P_{\rm abs}$) and radiated ($P_{\rm rad}$) powers in confined 2DES with current distribution ${\bf j}_\omega({\bf r})$. We start with power radiated by a small dipole, 
\begin{equation}
\label{Radiated_dipole}
    P_{\rm rad} = Z_0 \frac{\omega^4}{3\pi c^2}|{\bf d}_\omega|^2.
\end{equation}
We re-express the dipole moment via current distribution using its definition, ${\bf d}_\omega = \int{{\bf r} \rho_\omega({\bf r}) d^2 {\bf r}}$, and continuity equation $-i\omega \rho_\omega + \partial_{\bf r} {\bf j} =0$. This results in:
\begin{equation}
\label{Dipole_current}
    {\bf d}_\omega = \frac{1}{i\omega}\int{{\bf j}_\omega ({\bf r})d{\bf r}}.
\end{equation}
Introducing (\ref{Dipole_current}) into (\ref{Radiated_dipole}), and expressing the frequency via wavelength, $\omega = 2\pi c/\lambda_0$, we find:
\begin{equation}
  P_{\rm rad} = Z_0 \frac{4\pi }{3 \lambda_0^2}\left|\int{{\bf j}_\omega ({\bf r})d{\bf r}}\right|^2, 
\end{equation}
which is equivalent to the left-hand side of Eq.~(\ref{eq:match_isolated}) of the main text. As for absorbed power, it is simply given by Joule's law:
\begin{equation}
    P_{\rm abs} = 2 {\rm Re}[Z_{\rm 2des}(\omega)] \int{|{\bf j}_\omega({\bf r})|^2 d{\bf r}}.
\end{equation}
For Drude model of conductivity, ${\rm Re}[Z_{\rm 2des}(\omega)]$ is nothing but dc sheet resistance $R_\square $. This brings us exactly to Eq.~(\ref{eq:match_isolated}) of the main text.

\section{Analytical model for absorption of contacted 2d plasmonic device}
\label{app:reciprocity}
The developed model for resonant frequencies and matching conditions can be further extended to a fully analytical estimates of absorption cross-section. It is based on reciprocity theorem for antennas~\cite{balanis2015antenna}, and enables to present a cross-section of contacted plasmonic resonator as
\begin{equation} \label{eq:sigma}
    \sigma = \frac{\lambda_0^2}{4\pi} e_{\rm match} D(\theta, \phi),
\end{equation}
where $D(\theta,\phi)$ is the antenna directivity and 
\begin{equation}
e_{\rm match} = \frac{4 R_{\rm rad} R_{\rm 2des}}{|Z_A + Z^*_{\rm 2des}|^2}  
\end{equation}
is the matching efficiency which turns to unity for a perfectly-matched device. For dipole radiation, which is always the case for fundamental mode of sub-wavelength plasmonic resonator, $D(\theta,\phi) = 3/2 \sin^2\theta$, and $\theta$ is the angle between dipole direction and wave vector of incident (radiated) light. We  observe that the antenna model for cross-sections works very well for prediction of absorption cross-sections, as seen from comparison of dashed and solid lines in Fig.~\ref{fig3}a.

\section{A tunable resonant detector with low-mobility 2DES}
\label{app:spectrometer}
Another key advantage of 2D plasmonic resonators, apart from deep-sub-wavelength size, is the tuning of eigen-frequency upon variations of sheet conductivity $Z_{\rm 2des}$. The latter is enabled in situ by electric field effect or transverse magnetic field. Frequency tuning enables the realisation of all-electrical plasmonic radiation spectrometers. The deep confinement for such devices is unimportant, while large absorption cross-section is still highly desirable.

\begin{figure}[h] \centering
   \includegraphics[width=0.5\textwidth]{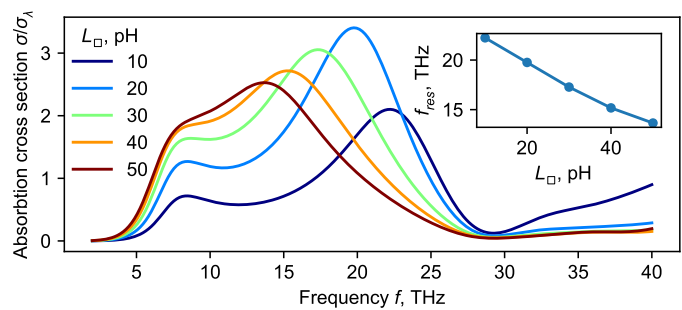}
  \caption{{\bf Resonant and tunable absorption by low-mobility 2DES.} Calculated spectrum of absorption cross section for small ($g=0.1$ $\mu$m) moderate-quality 2DES ($\tau=20$ fs) coupled to large antenna $L_a=16$ $\mu$m, $w=8$ $\mu$m. The highest peak occurs at frequency where $\lambda_0 \sim L_a$}
  \label{fig5}
\end{figure}

We show that a functional all-electric spectrometer (or tunable resonant detector) can be realised with 'dirty' 2d electron systems and resonant metal antennas with size $\sim \lambda_0$. Figure \ref{fig5} substantiates our findings by showing the absorption spectrum of 2DES-loaded antenna ($L_a = 16$ $\mu$m, $g=w=8$ $\mu$m) at various 2D inductances. Instructively, the cross-section at the second resonance reaches $(1...3)\sigma_\lambda$, while the momentum relaxation time is moderate, $\tau = 20$ fs. The resonant frequency goes down with increasing $L_\square$ (inset), as it typically occurs in plasmonic systems.

The origin of large cross section and wide frequency tuning in this example come from the special design of arms having the length order of $\lambda_0$. Electric current has a node in the middle of the structure, and reaches large values in the metal arms. This enables matching of high-impedance structures to free space. On the other hand, bare antenna impedance strongly varies with frequency once its length is close to $\lambda_0$, and small variations of load impedance can lead to pronounced variations of the net resonant frequency.

\bibliography{Sample}

\begin{thebibliography}{54}%
\makeatletter
\providecommand \@ifxundefined [1]{%
 \@ifx{#1\undefined}
}%
\providecommand \@ifnum [1]{%
 \ifnum #1\expandafter \@firstoftwo
 \else \expandafter \@secondoftwo
 \fi
}%
\providecommand \@ifx [1]{%
 \ifx #1\expandafter \@firstoftwo
 \else \expandafter \@secondoftwo
 \fi
}%
\providecommand \natexlab [1]{#1}%
\providecommand \enquote  [1]{``#1''}%
\providecommand \bibnamefont  [1]{#1}%
\providecommand \bibfnamefont [1]{#1}%
\providecommand \citenamefont [1]{#1}%
\providecommand \href@noop [0]{\@secondoftwo}%
\providecommand \href [0]{\begingroup \@sanitize@url \@href}%
\providecommand \@href[1]{\@@startlink{#1}\@@href}%
\providecommand \@@href[1]{\endgroup#1\@@endlink}%
\providecommand \@sanitize@url [0]{\catcode `\\12\catcode `\$12\catcode
  `\&12\catcode `\#12\catcode `\^12\catcode `\_12\catcode `\%12\relax}%
\providecommand \@@startlink[1]{}%
\providecommand \@@endlink[0]{}%
\providecommand \url  [0]{\begingroup\@sanitize@url \@url }%
\providecommand \@url [1]{\endgroup\@href {#1}{\urlprefix }}%
\providecommand \urlprefix  [0]{URL }%
\providecommand \Eprint [0]{\href }%
\providecommand \doibase [0]{https://doi.org/}%
\providecommand \selectlanguage [0]{\@gobble}%
\providecommand \bibinfo  [0]{\@secondoftwo}%
\providecommand \bibfield  [0]{\@secondoftwo}%
\providecommand \translation [1]{[#1]}%
\providecommand \BibitemOpen [0]{}%
\providecommand \bibitemStop [0]{}%
\providecommand \bibitemNoStop [0]{.\EOS\space}%
\providecommand \EOS [0]{\spacefactor3000\relax}%
\providecommand \BibitemShut  [1]{\csname bibitem#1\endcsname}%
\let\auto@bib@innerbib\@empty
\bibitem [{\citenamefont {Stern}(1967)}]{Stern1967}%
  \BibitemOpen
  \bibfield  {author} {\bibinfo {author} {\bibfnamefont {F.}~\bibnamefont
  {Stern}},\ }\bibfield  {title} {\bibinfo {title} {{Polarizability of a
  Two-Dimensional Electron Gas}},\ }\href
  {https://doi.org/10.1103/PhysRevLett.18.546} {\bibfield  {journal} {\bibinfo
  {journal} {Physical Review Letters}\ }\textbf {\bibinfo {volume} {18}},\
  \bibinfo {pages} {546} (\bibinfo {year} {1967})}\BibitemShut {NoStop}%
\bibitem [{\citenamefont {Koppens}\ \emph {et~al.}(2011)\citenamefont
  {Koppens}, \citenamefont {Chang},\ and\ \citenamefont {{Garc{\'{i}}a De
  Abajo}}}]{Koppens2011}%
  \BibitemOpen
  \bibfield  {author} {\bibinfo {author} {\bibfnamefont {F.~H.}\ \bibnamefont
  {Koppens}}, \bibinfo {author} {\bibfnamefont {D.~E.}\ \bibnamefont {Chang}},\
  and\ \bibinfo {author} {\bibfnamefont {F.~J.}\ \bibnamefont {{Garc{\'{i}}a De
  Abajo}}},\ }\bibfield  {title} {\bibinfo {title} {{Graphene plasmonics: A
  platform for strong light-matter interactions}},\ }\href
  {https://doi.org/10.1021/nl201771h} {\bibfield  {journal} {\bibinfo
  {journal} {Nano Letters}\ }\textbf {\bibinfo {volume} {11}},\ \bibinfo
  {pages} {3370} (\bibinfo {year} {2011})}\BibitemShut {NoStop}%
\bibitem [{\citenamefont {Grigorenko}\ \emph {et~al.}(2012)\citenamefont
  {Grigorenko}, \citenamefont {Polini},\ and\ \citenamefont
  {Novoselov}}]{Grigorenko2012}%
  \BibitemOpen
  \bibfield  {author} {\bibinfo {author} {\bibfnamefont {A.~N.}\ \bibnamefont
  {Grigorenko}}, \bibinfo {author} {\bibfnamefont {M.}~\bibnamefont {Polini}},\
  and\ \bibinfo {author} {\bibfnamefont {K.~S.}\ \bibnamefont {Novoselov}},\
  }\bibfield  {title} {\bibinfo {title} {{Graphene plasmonics}},\ }\href
  {https://doi.org/10.1038/nphoton.2012.262} {\bibfield  {journal} {\bibinfo
  {journal} {Nature Photonics}\ }\textbf {\bibinfo {volume} {6}},\ \bibinfo
  {pages} {749} (\bibinfo {year} {2012})},\ \Eprint
  {https://arxiv.org/abs/1301.4241} {arXiv:1301.4241} \BibitemShut {NoStop}%
\bibitem [{\citenamefont {García~de Abajo}(2014)}]{Abajo_GraphenePlasmonics}%
  \BibitemOpen
  \bibfield  {author} {\bibinfo {author} {\bibfnamefont {F.~J.}\ \bibnamefont
  {García~de Abajo}},\ }\bibfield  {title} {\bibinfo {title} {Graphene
  plasmonics: Challenges and opportunities},\ }\href
  {https://doi.org/10.1021/ph400147y} {\bibfield  {journal} {\bibinfo
  {journal} {ACS Photonics}\ }\textbf {\bibinfo {volume} {1}},\ \bibinfo
  {pages} {135} (\bibinfo {year} {2014})}\BibitemShut {NoStop}%
\bibitem [{\citenamefont {Ni}\ \emph {et~al.}(2018)\citenamefont {Ni},
  \citenamefont {McLeod}, \citenamefont {Sun}, \citenamefont {Wang},
  \citenamefont {Xiong}, \citenamefont {Post}, \citenamefont {Sunku},
  \citenamefont {Jiang}, \citenamefont {Hone}, \citenamefont {Dean},
  \citenamefont {Fogler},\ and\ \citenamefont
  {Basov}}]{Ni_LimitsToPlasmonics_2018}%
  \BibitemOpen
  \bibfield  {author} {\bibinfo {author} {\bibfnamefont {G.~X.}\ \bibnamefont
  {Ni}}, \bibinfo {author} {\bibfnamefont {A.~S.}\ \bibnamefont {McLeod}},
  \bibinfo {author} {\bibfnamefont {Z.}~\bibnamefont {Sun}}, \bibinfo {author}
  {\bibfnamefont {L.}~\bibnamefont {Wang}}, \bibinfo {author} {\bibfnamefont
  {L.}~\bibnamefont {Xiong}}, \bibinfo {author} {\bibfnamefont {K.~W.}\
  \bibnamefont {Post}}, \bibinfo {author} {\bibfnamefont {S.~S.}\ \bibnamefont
  {Sunku}}, \bibinfo {author} {\bibfnamefont {B.-Y.}\ \bibnamefont {Jiang}},
  \bibinfo {author} {\bibfnamefont {J.}~\bibnamefont {Hone}}, \bibinfo {author}
  {\bibfnamefont {C.~R.}\ \bibnamefont {Dean}}, \bibinfo {author}
  {\bibfnamefont {M.~M.}\ \bibnamefont {Fogler}},\ and\ \bibinfo {author}
  {\bibfnamefont {D.~N.}\ \bibnamefont {Basov}},\ }\bibfield  {title} {\bibinfo
  {title} {{Fundamental limits to graphene plasmonics}},\ }\href
  {https://doi.org/10.1038/s41586-018-0136-9} {\bibfield  {journal} {\bibinfo
  {journal} {Nature}\ }\textbf {\bibinfo {volume} {557}},\ \bibinfo {pages}
  {530} (\bibinfo {year} {2018})}\BibitemShut {NoStop}%
\bibitem [{\citenamefont {Woessner}\ \emph {et~al.}(2015)\citenamefont
  {Woessner}, \citenamefont {Lundeberg}, \citenamefont {Gao}, \citenamefont
  {Principi}, \citenamefont {Alonso-Gonz{\'{a}}lez}, \citenamefont {Carrega},
  \citenamefont {Watanabe}, \citenamefont {Taniguchi}, \citenamefont {Vignale},
  \citenamefont {Polini}, \citenamefont {Hone}, \citenamefont {Hillenbrand},\
  and\ \citenamefont {Koppens}}]{Woessner2015}%
  \BibitemOpen
  \bibfield  {author} {\bibinfo {author} {\bibfnamefont {A.}~\bibnamefont
  {Woessner}}, \bibinfo {author} {\bibfnamefont {M.~B.}\ \bibnamefont
  {Lundeberg}}, \bibinfo {author} {\bibfnamefont {Y.}~\bibnamefont {Gao}},
  \bibinfo {author} {\bibfnamefont {A.}~\bibnamefont {Principi}}, \bibinfo
  {author} {\bibfnamefont {P.}~\bibnamefont {Alonso-Gonz{\'{a}}lez}}, \bibinfo
  {author} {\bibfnamefont {M.}~\bibnamefont {Carrega}}, \bibinfo {author}
  {\bibfnamefont {K.}~\bibnamefont {Watanabe}}, \bibinfo {author}
  {\bibfnamefont {T.}~\bibnamefont {Taniguchi}}, \bibinfo {author}
  {\bibfnamefont {G.}~\bibnamefont {Vignale}}, \bibinfo {author} {\bibfnamefont
  {M.}~\bibnamefont {Polini}}, \bibinfo {author} {\bibfnamefont
  {J.}~\bibnamefont {Hone}}, \bibinfo {author} {\bibfnamefont {R.}~\bibnamefont
  {Hillenbrand}},\ and\ \bibinfo {author} {\bibfnamefont {F.~H.~L.}\
  \bibnamefont {Koppens}},\ }\bibfield  {title} {\bibinfo {title} {{Highly
  confined low-loss plasmons in graphene–boron nitride heterostructures}},\
  }\href {https://doi.org/10.1038/nmat4169} {\bibfield  {journal} {\bibinfo
  {journal} {Nature Materials}\ }\textbf {\bibinfo {volume} {14}},\ \bibinfo
  {pages} {421} (\bibinfo {year} {2015})},\ \Eprint
  {https://arxiv.org/abs/1409.5674} {1409.5674} \BibitemShut {NoStop}%
\bibitem [{\citenamefont {Scalari}\ \emph {et~al.}(2012)\citenamefont
  {Scalari}, \citenamefont {Maissen}, \citenamefont {Turcinkova}, \citenamefont
  {Hagenmuller}, \citenamefont {{De Liberato}}, \citenamefont {Ciuti},
  \citenamefont {Reichl}, \citenamefont {Schuh}, \citenamefont {Wegscheider},
  \citenamefont {Beck},\ and\ \citenamefont
  {Faist}}]{Faist_UltraStrongCoupling}%
  \BibitemOpen
  \bibfield  {author} {\bibinfo {author} {\bibfnamefont {G.}~\bibnamefont
  {Scalari}}, \bibinfo {author} {\bibfnamefont {C.}~\bibnamefont {Maissen}},
  \bibinfo {author} {\bibfnamefont {D.}~\bibnamefont {Turcinkova}}, \bibinfo
  {author} {\bibfnamefont {D.}~\bibnamefont {Hagenmuller}}, \bibinfo {author}
  {\bibfnamefont {S.}~\bibnamefont {{De Liberato}}}, \bibinfo {author}
  {\bibfnamefont {C.}~\bibnamefont {Ciuti}}, \bibinfo {author} {\bibfnamefont
  {C.}~\bibnamefont {Reichl}}, \bibinfo {author} {\bibfnamefont
  {D.}~\bibnamefont {Schuh}}, \bibinfo {author} {\bibfnamefont
  {W.}~\bibnamefont {Wegscheider}}, \bibinfo {author} {\bibfnamefont
  {M.}~\bibnamefont {Beck}},\ and\ \bibinfo {author} {\bibfnamefont
  {J.}~\bibnamefont {Faist}},\ }\bibfield  {title} {\bibinfo {title}
  {{Ultrastrong Coupling of the Cyclotron Transition of a 2D Electron Gas to a
  THz Metamaterial}},\ }\href {https://doi.org/10.1126/science.1216022}
  {\bibfield  {journal} {\bibinfo  {journal} {Science}\ }\textbf {\bibinfo
  {volume} {335}},\ \bibinfo {pages} {1323} (\bibinfo {year}
  {2012})}\BibitemShut {NoStop}%
\bibitem [{\citenamefont {Alonso-Gonzalez}\ \emph {et~al.}(2014)\citenamefont
  {Alonso-Gonzalez}, \citenamefont {Nikitin}, \citenamefont {Golmar},
  \citenamefont {Centeno}, \citenamefont {Pesquera}, \citenamefont {Velez},
  \citenamefont {Chen}, \citenamefont {Navickaite}, \citenamefont {Koppens},
  \citenamefont {Zurutuza}, \citenamefont {Casanova}, \citenamefont {Hueso},\
  and\ \citenamefont {Hillenbrand}}]{Alonso-Gonzalez2014}%
  \BibitemOpen
  \bibfield  {author} {\bibinfo {author} {\bibfnamefont {P.}~\bibnamefont
  {Alonso-Gonzalez}}, \bibinfo {author} {\bibfnamefont {A.~Y.}\ \bibnamefont
  {Nikitin}}, \bibinfo {author} {\bibfnamefont {F.}~\bibnamefont {Golmar}},
  \bibinfo {author} {\bibfnamefont {A.}~\bibnamefont {Centeno}}, \bibinfo
  {author} {\bibfnamefont {A.}~\bibnamefont {Pesquera}}, \bibinfo {author}
  {\bibfnamefont {S.}~\bibnamefont {Velez}}, \bibinfo {author} {\bibfnamefont
  {J.}~\bibnamefont {Chen}}, \bibinfo {author} {\bibfnamefont {G.}~\bibnamefont
  {Navickaite}}, \bibinfo {author} {\bibfnamefont {F.}~\bibnamefont {Koppens}},
  \bibinfo {author} {\bibfnamefont {A.}~\bibnamefont {Zurutuza}}, \bibinfo
  {author} {\bibfnamefont {F.}~\bibnamefont {Casanova}}, \bibinfo {author}
  {\bibfnamefont {L.~E.}\ \bibnamefont {Hueso}},\ and\ \bibinfo {author}
  {\bibfnamefont {R.}~\bibnamefont {Hillenbrand}},\ }\bibfield  {title}
  {\bibinfo {title} {{Controlling graphene plasmons with resonant metal
  antennas and spatial conductivity patterns}},\ }\href
  {https://doi.org/10.1126/science.1253202} {\bibfield  {journal} {\bibinfo
  {journal} {Science}\ }\textbf {\bibinfo {volume} {344}},\ \bibinfo {pages}
  {1369} (\bibinfo {year} {2014})}\BibitemShut {NoStop}%
\bibitem [{\citenamefont {Dyakonov}\ and\ \citenamefont
  {Shur}(1996)}]{Dyakonov1996}%
  \BibitemOpen
  \bibfield  {author} {\bibinfo {author} {\bibfnamefont {M.}~\bibnamefont
  {Dyakonov}}\ and\ \bibinfo {author} {\bibfnamefont {M.}~\bibnamefont
  {Shur}},\ }\bibfield  {title} {\bibinfo {title} {Detection, mixing, and
  frequency multiplication of terahertz radiation by two-dimensional electronic
  fluid},\ }\href {https://doi.org/10.1109/16.485650} {\bibfield  {journal}
  {\bibinfo  {journal} {IEEE Transactions on Electron Devices}\ }\textbf
  {\bibinfo {volume} {43}},\ \bibinfo {pages} {380} (\bibinfo {year}
  {1996})}\BibitemShut {NoStop}%
\bibitem [{\citenamefont {Atwater}\ and\ \citenamefont
  {Polman}(2010)}]{Atwater2010}%
  \BibitemOpen
  \bibfield  {author} {\bibinfo {author} {\bibfnamefont {H.~A.}\ \bibnamefont
  {Atwater}}\ and\ \bibinfo {author} {\bibfnamefont {A.}~\bibnamefont
  {Polman}},\ }\bibfield  {title} {\bibinfo {title} {Plasmonics for improved
  photovoltaic devices},\ }\href {https://doi.org/10.1038/nmat2866} {\bibfield
  {journal} {\bibinfo  {journal} {Nature Materials}\ }\textbf {\bibinfo
  {volume} {9}},\ \bibinfo {pages} {865} (\bibinfo {year} {2010})}\BibitemShut
  {NoStop}%
\bibitem [{\citenamefont {Sengupta}\ \emph {et~al.}(2018)\citenamefont
  {Sengupta}, \citenamefont {Nagatsuma},\ and\ \citenamefont
  {Mittleman}}]{MIttleman_NatEl_THz}%
  \BibitemOpen
  \bibfield  {author} {\bibinfo {author} {\bibfnamefont {K.}~\bibnamefont
  {Sengupta}}, \bibinfo {author} {\bibfnamefont {T.}~\bibnamefont
  {Nagatsuma}},\ and\ \bibinfo {author} {\bibfnamefont {D.~M.}\ \bibnamefont
  {Mittleman}},\ }\bibfield  {title} {\bibinfo {title} {{Terahertz integrated
  electronic and hybrid electronic–photonic systems}},\ }\href
  {https://doi.org/10.1038/s41928-018-0173-2} {\bibfield  {journal} {\bibinfo
  {journal} {Nature Electronics}\ }\textbf {\bibinfo {volume} {1}},\ \bibinfo
  {pages} {622} (\bibinfo {year} {2018})}\BibitemShut {NoStop}%
\bibitem [{\citenamefont {Tredicucci}\ and\ \citenamefont
  {Vitiello}(2014)}]{Tredicucci_DeviceConceptsGraphene}%
  \BibitemOpen
  \bibfield  {author} {\bibinfo {author} {\bibfnamefont {A.}~\bibnamefont
  {Tredicucci}}\ and\ \bibinfo {author} {\bibfnamefont {M.~S.}\ \bibnamefont
  {Vitiello}},\ }\bibfield  {title} {\bibinfo {title} {Device concepts for
  graphene-based terahertz photonics},\ }\href
  {https://doi.org/10.1109/JSTQE.2013.2271692} {\bibfield  {journal} {\bibinfo
  {journal} {IEEE Journal of Selected Topics in Quantum Electronics}\ }\textbf
  {\bibinfo {volume} {20}},\ \bibinfo {pages} {130} (\bibinfo {year}
  {2014})}\BibitemShut {NoStop}%
\bibitem [{\citenamefont {Viti}\ \emph {et~al.}(2016)\citenamefont {Viti},
  \citenamefont {Coquillat}, \citenamefont {Politano}, \citenamefont {Kokh},
  \citenamefont {Aliev}, \citenamefont {Babanly}, \citenamefont {Tereshchenko},
  \citenamefont {Knap}, \citenamefont {Chulkov},\ and\ \citenamefont
  {Vitiello}}]{Viti_Detection_TI_Surface_states}%
  \BibitemOpen
  \bibfield  {author} {\bibinfo {author} {\bibfnamefont {L.}~\bibnamefont
  {Viti}}, \bibinfo {author} {\bibfnamefont {D.}~\bibnamefont {Coquillat}},
  \bibinfo {author} {\bibfnamefont {A.}~\bibnamefont {Politano}}, \bibinfo
  {author} {\bibfnamefont {K.~A.}\ \bibnamefont {Kokh}}, \bibinfo {author}
  {\bibfnamefont {Z.~S.}\ \bibnamefont {Aliev}}, \bibinfo {author}
  {\bibfnamefont {M.~B.}\ \bibnamefont {Babanly}}, \bibinfo {author}
  {\bibfnamefont {O.~E.}\ \bibnamefont {Tereshchenko}}, \bibinfo {author}
  {\bibfnamefont {W.}~\bibnamefont {Knap}}, \bibinfo {author} {\bibfnamefont
  {E.~V.}\ \bibnamefont {Chulkov}},\ and\ \bibinfo {author} {\bibfnamefont
  {M.~S.}\ \bibnamefont {Vitiello}},\ }\bibfield  {title} {\bibinfo {title}
  {Plasma-wave terahertz detection mediated by topological insulators surface
  states},\ }\href {https://doi.org/10.1021/acs.nanolett.5b02901} {\bibfield
  {journal} {\bibinfo  {journal} {Nano Letters}\ }\textbf {\bibinfo {volume}
  {16}},\ \bibinfo {pages} {80} (\bibinfo {year} {2016})}\BibitemShut {NoStop}%
\bibitem [{\citenamefont {Viti}\ \emph {et~al.}(2015)\citenamefont {Viti},
  \citenamefont {Hu}, \citenamefont {Coquillat}, \citenamefont {Knap},
  \citenamefont {Tredicucci}, \citenamefont {Politano},\ and\ \citenamefont
  {Vitiello}}]{Viti_BlackP}%
  \BibitemOpen
  \bibfield  {author} {\bibinfo {author} {\bibfnamefont {L.}~\bibnamefont
  {Viti}}, \bibinfo {author} {\bibfnamefont {J.}~\bibnamefont {Hu}}, \bibinfo
  {author} {\bibfnamefont {D.}~\bibnamefont {Coquillat}}, \bibinfo {author}
  {\bibfnamefont {W.}~\bibnamefont {Knap}}, \bibinfo {author} {\bibfnamefont
  {A.}~\bibnamefont {Tredicucci}}, \bibinfo {author} {\bibfnamefont
  {A.}~\bibnamefont {Politano}},\ and\ \bibinfo {author} {\bibfnamefont
  {M.~S.}\ \bibnamefont {Vitiello}},\ }\bibfield  {title} {\bibinfo {title}
  {Black phosphorus terahertz photodetectors},\ }\href
  {https://doi.org/https://doi.org/10.1002/adma.201502052} {\bibfield
  {journal} {\bibinfo  {journal} {Advanced Materials}\ }\textbf {\bibinfo
  {volume} {27}},\ \bibinfo {pages} {5567} (\bibinfo {year}
  {2015})}\BibitemShut {NoStop}%
\bibitem [{\citenamefont {Yavorskiy}\ \emph {et~al.}(2020)\citenamefont
  {Yavorskiy}, \citenamefont {Szoła}, \citenamefont {Karpierz}, \citenamefont
  {Bożek}, \citenamefont {Rudniewski}, \citenamefont {Karczewski},
  \citenamefont {Wojtowicz}, \citenamefont {Wróbel},\ and\ \citenamefont
  {Łusakowski}}]{Yavorsky_HgTe_detection}%
  \BibitemOpen
  \bibfield  {author} {\bibinfo {author} {\bibfnamefont {D.}~\bibnamefont
  {Yavorskiy}}, \bibinfo {author} {\bibfnamefont {M.}~\bibnamefont {Szoła}},
  \bibinfo {author} {\bibfnamefont {K.}~\bibnamefont {Karpierz}}, \bibinfo
  {author} {\bibfnamefont {R.}~\bibnamefont {Bożek}}, \bibinfo {author}
  {\bibfnamefont {R.}~\bibnamefont {Rudniewski}}, \bibinfo {author}
  {\bibfnamefont {G.}~\bibnamefont {Karczewski}}, \bibinfo {author}
  {\bibfnamefont {T.}~\bibnamefont {Wojtowicz}}, \bibinfo {author}
  {\bibfnamefont {J.}~\bibnamefont {Wróbel}},\ and\ \bibinfo {author}
  {\bibfnamefont {J.}~\bibnamefont {Łusakowski}},\ }\bibfield  {title}
  {\bibinfo {title} {Grating metamaterials based on cdte/cdmgte quantum wells
  as terahertz detectors for high magnetic field applications},\ }\bibfield
  {journal} {\bibinfo  {journal} {Applied Sciences}\ }\textbf {\bibinfo
  {volume} {10}},\ \href {https://doi.org/10.3390/app10082807}
  {10.3390/app10082807} (\bibinfo {year} {2020})\BibitemShut {NoStop}%
\bibitem [{\citenamefont {Tomadin}\ and\ \citenamefont
  {Polini}(2013)}]{Tomadin_PRB_PlasmawaveResponse}%
  \BibitemOpen
  \bibfield  {author} {\bibinfo {author} {\bibfnamefont {A.}~\bibnamefont
  {Tomadin}}\ and\ \bibinfo {author} {\bibfnamefont {M.}~\bibnamefont
  {Polini}},\ }\bibfield  {title} {\bibinfo {title} {Theory of the plasma-wave
  photoresponse of a gated graphene sheet},\ }\href
  {https://doi.org/10.1103/PhysRevB.88.205426} {\bibfield  {journal} {\bibinfo
  {journal} {Phys. Rev. B}\ }\textbf {\bibinfo {volume} {88}},\ \bibinfo
  {pages} {205426} (\bibinfo {year} {2013})}\BibitemShut {NoStop}%
\bibitem [{\citenamefont {Winstanley}\ \emph {et~al.}(2021)\citenamefont
  {Winstanley}, \citenamefont {Schomerus},\ and\ \citenamefont
  {Principi}}]{Principi_PRB_CorbinoDetector}%
  \BibitemOpen
  \bibfield  {author} {\bibinfo {author} {\bibfnamefont {B.}~\bibnamefont
  {Winstanley}}, \bibinfo {author} {\bibfnamefont {H.}~\bibnamefont
  {Schomerus}},\ and\ \bibinfo {author} {\bibfnamefont {A.}~\bibnamefont
  {Principi}},\ }\bibfield  {title} {\bibinfo {title} {Corbino field-effect
  transistors in a magnetic field: Highly tunable photodetectors},\ }\href
  {https://doi.org/10.1103/PhysRevB.104.165406} {\bibfield  {journal} {\bibinfo
   {journal} {Phys. Rev. B}\ }\textbf {\bibinfo {volume} {104}},\ \bibinfo
  {pages} {165406} (\bibinfo {year} {2021})}\BibitemShut {NoStop}%
\bibitem [{\citenamefont {Ryzhii}\ and\ \citenamefont
  {Shur}(2006)}]{Ryzhii_Shur_JJAP}%
  \BibitemOpen
  \bibfield  {author} {\bibinfo {author} {\bibfnamefont {V.}~\bibnamefont
  {Ryzhii}}\ and\ \bibinfo {author} {\bibfnamefont {M.~S.}\ \bibnamefont
  {Shur}},\ }\bibfield  {title} {\bibinfo {title} {Resonant terahertz detector
  utilizing plasma oscillations in two-dimensional electron system with lateral
  schottky junction},\ }\href {https://doi.org/10.1143/jjap.45.l1118}
  {\bibfield  {journal} {\bibinfo  {journal} {Japanese Journal of Applied
  Physics}\ }\textbf {\bibinfo {volume} {45}},\ \bibinfo {pages} {L1118}
  (\bibinfo {year} {2006})}\BibitemShut {NoStop}%
\bibitem [{\citenamefont {Popov}(2013)}]{Popov_rectification_plasma_waves}%
  \BibitemOpen
  \bibfield  {author} {\bibinfo {author} {\bibfnamefont {V.~V.}\ \bibnamefont
  {Popov}},\ }\bibfield  {title} {\bibinfo {title} {Terahertz rectification by
  periodic two-dimensional electron plasma},\ }\href
  {https://doi.org/10.1063/1.4811706} {\bibfield  {journal} {\bibinfo
  {journal} {Applied Physics Letters}\ }\textbf {\bibinfo {volume} {102}},\
  \bibinfo {pages} {253504} (\bibinfo {year} {2013})}\BibitemShut {NoStop}%
\bibitem [{\citenamefont {Principi}\ \emph {et~al.}(2019)\citenamefont
  {Principi}, \citenamefont {Bandurin}, \citenamefont {Rostami},\ and\
  \citenamefont {Polini}}]{Pseudo_Euler}%
  \BibitemOpen
  \bibfield  {author} {\bibinfo {author} {\bibfnamefont {A.}~\bibnamefont
  {Principi}}, \bibinfo {author} {\bibfnamefont {D.}~\bibnamefont {Bandurin}},
  \bibinfo {author} {\bibfnamefont {H.}~\bibnamefont {Rostami}},\ and\ \bibinfo
  {author} {\bibfnamefont {M.}~\bibnamefont {Polini}},\ }\bibfield  {title}
  {\bibinfo {title} {Pseudo-euler equations from nonlinear optics:
  Plasmon-assisted photodetection beyond hydrodynamics},\ }\href
  {https://doi.org/10.1103/PhysRevB.99.075410} {\bibfield  {journal} {\bibinfo
  {journal} {Phys. Rev. B}\ }\textbf {\bibinfo {volume} {99}},\ \bibinfo
  {pages} {075410} (\bibinfo {year} {2019})}\BibitemShut {NoStop}%
\bibitem [{\citenamefont {Kachorovskii}\ \emph {et~al.}(2013)\citenamefont
  {Kachorovskii}, \citenamefont {Rumyantsev}, \citenamefont {Knap},\ and\
  \citenamefont {Shur}}]{Kachorovskii_performance_limits}%
  \BibitemOpen
  \bibfield  {author} {\bibinfo {author} {\bibfnamefont {V.~Y.}\ \bibnamefont
  {Kachorovskii}}, \bibinfo {author} {\bibfnamefont {S.~L.}\ \bibnamefont
  {Rumyantsev}}, \bibinfo {author} {\bibfnamefont {W.}~\bibnamefont {Knap}},\
  and\ \bibinfo {author} {\bibfnamefont {M.}~\bibnamefont {Shur}},\ }\bibfield
  {title} {\bibinfo {title} {Performance limits for field effect transistors as
  terahertz detectors},\ }\href {https://doi.org/10.1063/1.4809672} {\bibfield
  {journal} {\bibinfo  {journal} {Applied Physics Letters}\ }\textbf {\bibinfo
  {volume} {102}},\ \bibinfo {pages} {223505} (\bibinfo {year}
  {2013})}\BibitemShut {NoStop}%
\bibitem [{\citenamefont {Olbrich}\ \emph {et~al.}(2016)\citenamefont
  {Olbrich}, \citenamefont {Kamann}, \citenamefont {K{\"{o}}nig}, \citenamefont
  {Munzert}, \citenamefont {Tutsch}, \citenamefont {Eroms}, \citenamefont
  {Weiss}, \citenamefont {Liu}, \citenamefont {Golub}, \citenamefont
  {Ivchenko}, \citenamefont {Popov}, \citenamefont {Fateev}, \citenamefont
  {Mashinsky}, \citenamefont {Fromm}, \citenamefont {Seyller},\ and\
  \citenamefont {Ganichev}}]{Olbrich2016}%
  \BibitemOpen
  \bibfield  {author} {\bibinfo {author} {\bibfnamefont {P.}~\bibnamefont
  {Olbrich}}, \bibinfo {author} {\bibfnamefont {J.}~\bibnamefont {Kamann}},
  \bibinfo {author} {\bibfnamefont {M.}~\bibnamefont {K{\"{o}}nig}}, \bibinfo
  {author} {\bibfnamefont {J.}~\bibnamefont {Munzert}}, \bibinfo {author}
  {\bibfnamefont {L.}~\bibnamefont {Tutsch}}, \bibinfo {author} {\bibfnamefont
  {J.}~\bibnamefont {Eroms}}, \bibinfo {author} {\bibfnamefont
  {D.}~\bibnamefont {Weiss}}, \bibinfo {author} {\bibfnamefont {M.-H.}\
  \bibnamefont {Liu}}, \bibinfo {author} {\bibfnamefont {L.~E.}\ \bibnamefont
  {Golub}}, \bibinfo {author} {\bibfnamefont {E.~L.}\ \bibnamefont {Ivchenko}},
  \bibinfo {author} {\bibfnamefont {V.~V.}\ \bibnamefont {Popov}}, \bibinfo
  {author} {\bibfnamefont {D.~V.}\ \bibnamefont {Fateev}}, \bibinfo {author}
  {\bibfnamefont {K.~V.}\ \bibnamefont {Mashinsky}}, \bibinfo {author}
  {\bibfnamefont {F.}~\bibnamefont {Fromm}}, \bibinfo {author} {\bibfnamefont
  {T.}~\bibnamefont {Seyller}},\ and\ \bibinfo {author} {\bibfnamefont {S.~D.}\
  \bibnamefont {Ganichev}},\ }\bibfield  {title} {\bibinfo {title} {{Terahertz
  ratchet effects in graphene with a lateral superlattice}},\ }\href
  {https://doi.org/10.1103/PhysRevB.93.075422} {\bibfield  {journal} {\bibinfo
  {journal} {Physical Review B}\ }\textbf {\bibinfo {volume} {93}},\ \bibinfo
  {pages} {075422} (\bibinfo {year} {2016})},\ \Eprint
  {https://arxiv.org/abs/1510.07946} {1510.07946} \BibitemShut {NoStop}%
\bibitem [{\citenamefont {Delgado-Notario}\ \emph {et~al.}(2020)\citenamefont
  {Delgado-Notario}, \citenamefont {Clericò}, \citenamefont {Diez},
  \citenamefont {Velázquez-Pérez}, \citenamefont {Taniguchi}, \citenamefont
  {Watanabe}, \citenamefont {Otsuji},\ and\ \citenamefont
  {Meziani}}]{Meziani_ADGG_graphene}%
  \BibitemOpen
  \bibfield  {author} {\bibinfo {author} {\bibfnamefont {J.~A.}\ \bibnamefont
  {Delgado-Notario}}, \bibinfo {author} {\bibfnamefont {V.}~\bibnamefont
  {Clericò}}, \bibinfo {author} {\bibfnamefont {E.}~\bibnamefont {Diez}},
  \bibinfo {author} {\bibfnamefont {J.~E.}\ \bibnamefont {Velázquez-Pérez}},
  \bibinfo {author} {\bibfnamefont {T.}~\bibnamefont {Taniguchi}}, \bibinfo
  {author} {\bibfnamefont {K.}~\bibnamefont {Watanabe}}, \bibinfo {author}
  {\bibfnamefont {T.}~\bibnamefont {Otsuji}},\ and\ \bibinfo {author}
  {\bibfnamefont {Y.~M.}\ \bibnamefont {Meziani}},\ }\bibfield  {title}
  {\bibinfo {title} {Asymmetric dual-grating gates graphene fet for detection
  of terahertz radiations},\ }\href {https://doi.org/10.1063/5.0007249}
  {\bibfield  {journal} {\bibinfo  {journal} {APL Photonics}\ }\textbf
  {\bibinfo {volume} {5}},\ \bibinfo {pages} {066102} (\bibinfo {year}
  {2020})}\BibitemShut {NoStop}%
\bibitem [{\citenamefont {Dorozhkin}\ \emph {et~al.}(2005)\citenamefont
  {Dorozhkin}, \citenamefont {Tovstonog}, \citenamefont {Mikhailov},
  \citenamefont {Kukushkin}, \citenamefont {Smet}, \citenamefont {{Von
  Klitzing}}, \citenamefont {Klitzing},\ and\ \citenamefont {{Von
  Klitzing}}}]{Dorozhkin2005}%
  \BibitemOpen
  \bibfield  {author} {\bibinfo {author} {\bibfnamefont {P.~S.}\ \bibnamefont
  {Dorozhkin}}, \bibinfo {author} {\bibfnamefont {S.~V.}\ \bibnamefont
  {Tovstonog}}, \bibinfo {author} {\bibfnamefont {S.~A.}\ \bibnamefont
  {Mikhailov}}, \bibinfo {author} {\bibfnamefont {I.~V.}\ \bibnamefont
  {Kukushkin}}, \bibinfo {author} {\bibfnamefont {J.~H.}\ \bibnamefont {Smet}},
  \bibinfo {author} {\bibfnamefont {K.}~\bibnamefont {{Von Klitzing}}},
  \bibinfo {author} {\bibfnamefont {K.~V.}\ \bibnamefont {Klitzing}},\ and\
  \bibinfo {author} {\bibfnamefont {K.}~\bibnamefont {{Von Klitzing}}},\
  }\bibfield  {title} {\bibinfo {title} {{Resonant detection of microwave
  radiation in a circular two-dimensional electron system with quantum point
  contacts}},\ }\href {https://doi.org/10.1063/1.2035883} {\bibfield  {journal}
  {\bibinfo  {journal} {Applied Physics Letters}\ }\textbf {\bibinfo {volume}
  {87}},\ \bibinfo {pages} {5} (\bibinfo {year} {2005})}\BibitemShut {NoStop}%
\bibitem [{\citenamefont {Muravev}\ and\ \citenamefont
  {Kukushkin}(2012)}]{Muravev2012a}%
  \BibitemOpen
  \bibfield  {author} {\bibinfo {author} {\bibfnamefont {V.~M.}\ \bibnamefont
  {Muravev}}\ and\ \bibinfo {author} {\bibfnamefont {I.~V.}\ \bibnamefont
  {Kukushkin}},\ }\bibfield  {title} {\bibinfo {title} {{Plasmonic
  detector/spectrometer of subterahertz radiation based on two-dimensional
  electron system with embedded defect}},\ }\href
  {https://doi.org/10.1063/1.3688049} {\bibfield  {journal} {\bibinfo
  {journal} {Applied Physics Letters}\ }\textbf {\bibinfo {volume} {100}},\
  \bibinfo {pages} {10} (\bibinfo {year} {2012})}\BibitemShut {NoStop}%
\bibitem [{\citenamefont {Knap}\ \emph {et~al.}(2002)\citenamefont {Knap},
  \citenamefont {Deng}, \citenamefont {Rumyantsev}, \citenamefont {Lü},
  \citenamefont {Shur}, \citenamefont {Saylor},\ and\ \citenamefont
  {Brunel}}]{Knap_resonant}%
  \BibitemOpen
  \bibfield  {author} {\bibinfo {author} {\bibfnamefont {W.}~\bibnamefont
  {Knap}}, \bibinfo {author} {\bibfnamefont {Y.}~\bibnamefont {Deng}}, \bibinfo
  {author} {\bibfnamefont {S.}~\bibnamefont {Rumyantsev}}, \bibinfo {author}
  {\bibfnamefont {J.-Q.}\ \bibnamefont {Lü}}, \bibinfo {author} {\bibfnamefont
  {M.~S.}\ \bibnamefont {Shur}}, \bibinfo {author} {\bibfnamefont {C.~A.}\
  \bibnamefont {Saylor}},\ and\ \bibinfo {author} {\bibfnamefont {L.~C.}\
  \bibnamefont {Brunel}},\ }\bibfield  {title} {\bibinfo {title} {Resonant
  detection of subterahertz radiation by plasma waves in a submicron
  field-effect transistor},\ }\href {https://doi.org/10.1063/1.1473685}
  {\bibfield  {journal} {\bibinfo  {journal} {Applied Physics Letters}\
  }\textbf {\bibinfo {volume} {80}},\ \bibinfo {pages} {3433} (\bibinfo {year}
  {2002})}\BibitemShut {NoStop}%
\bibitem [{\citenamefont {Peralta}\ \emph {et~al.}(2002)\citenamefont
  {Peralta}, \citenamefont {Allen}, \citenamefont {Wanke}, \citenamefont
  {Harff}, \citenamefont {Simmons}, \citenamefont {Lilly}, \citenamefont
  {Reno}, \citenamefont {Burke},\ and\ \citenamefont
  {Eisenstein}}]{Peralta2002}%
  \BibitemOpen
  \bibfield  {author} {\bibinfo {author} {\bibfnamefont {X.~G.}\ \bibnamefont
  {Peralta}}, \bibinfo {author} {\bibfnamefont {S.~J.}\ \bibnamefont {Allen}},
  \bibinfo {author} {\bibfnamefont {M.~C.}\ \bibnamefont {Wanke}}, \bibinfo
  {author} {\bibfnamefont {N.~E.}\ \bibnamefont {Harff}}, \bibinfo {author}
  {\bibfnamefont {J.~A.}\ \bibnamefont {Simmons}}, \bibinfo {author}
  {\bibfnamefont {M.~P.}\ \bibnamefont {Lilly}}, \bibinfo {author}
  {\bibfnamefont {J.~L.}\ \bibnamefont {Reno}}, \bibinfo {author}
  {\bibfnamefont {P.~J.}\ \bibnamefont {Burke}},\ and\ \bibinfo {author}
  {\bibfnamefont {J.~P.}\ \bibnamefont {Eisenstein}},\ }\bibfield  {title}
  {\bibinfo {title} {{Terahertz photoconductivity and plasmon modes in
  double-quantum-well field-effect transistors}},\ }\href
  {https://doi.org/10.1063/1.1497433} {\bibfield  {journal} {\bibinfo
  {journal} {Applied Physics Letters}\ }\textbf {\bibinfo {volume} {81}},\
  \bibinfo {pages} {1627} (\bibinfo {year} {2002})}\BibitemShut {NoStop}%
\bibitem [{\citenamefont {Cai}\ \emph {et~al.}(2015)\citenamefont {Cai},
  \citenamefont {Sushkov}, \citenamefont {Jadidi}, \citenamefont {Nyakiti},
  \citenamefont {Myers-Ward}, \citenamefont {Gaskill}, \citenamefont {Murphy},
  \citenamefont {Fuhrer},\ and\ \citenamefont
  {Drew}}]{Cai_photocurrent_by_plasmons}%
  \BibitemOpen
  \bibfield  {author} {\bibinfo {author} {\bibfnamefont {X.}~\bibnamefont
  {Cai}}, \bibinfo {author} {\bibfnamefont {A.~B.}\ \bibnamefont {Sushkov}},
  \bibinfo {author} {\bibfnamefont {M.~M.}\ \bibnamefont {Jadidi}}, \bibinfo
  {author} {\bibfnamefont {L.~O.}\ \bibnamefont {Nyakiti}}, \bibinfo {author}
  {\bibfnamefont {R.~L.}\ \bibnamefont {Myers-Ward}}, \bibinfo {author}
  {\bibfnamefont {D.~K.}\ \bibnamefont {Gaskill}}, \bibinfo {author}
  {\bibfnamefont {T.~E.}\ \bibnamefont {Murphy}}, \bibinfo {author}
  {\bibfnamefont {M.~S.}\ \bibnamefont {Fuhrer}},\ and\ \bibinfo {author}
  {\bibfnamefont {H.~D.}\ \bibnamefont {Drew}},\ }\bibfield  {title} {\bibinfo
  {title} {{Plasmon-Enhanced Terahertz Photodetection in Graphene}},\ }\href
  {https://doi.org/10.1021/acs.nanolett.5b00137} {\bibfield  {journal}
  {\bibinfo  {journal} {Nano Letters}\ }\textbf {\bibinfo {volume} {15}},\
  \bibinfo {pages} {4295} (\bibinfo {year} {2015})}\BibitemShut {NoStop}%
\bibitem [{\citenamefont {Freitag}\ \emph {et~al.}(2013)\citenamefont
  {Freitag}, \citenamefont {Low}, \citenamefont {Zhu}, \citenamefont {Yan},
  \citenamefont {Xia},\ and\ \citenamefont
  {Avouris}}]{Freitag_Intrinsic_plasmons}%
  \BibitemOpen
  \bibfield  {author} {\bibinfo {author} {\bibfnamefont {M.}~\bibnamefont
  {Freitag}}, \bibinfo {author} {\bibfnamefont {T.}~\bibnamefont {Low}},
  \bibinfo {author} {\bibfnamefont {W.}~\bibnamefont {Zhu}}, \bibinfo {author}
  {\bibfnamefont {H.}~\bibnamefont {Yan}}, \bibinfo {author} {\bibfnamefont
  {F.}~\bibnamefont {Xia}},\ and\ \bibinfo {author} {\bibfnamefont
  {P.}~\bibnamefont {Avouris}},\ }\bibfield  {title} {\bibinfo {title}
  {{Photocurrent in graphene harnessed by tunable intrinsic plasmons}},\ }\href
  {https://doi.org/10.1038/ncomms2951} {\bibfield  {journal} {\bibinfo
  {journal} {Nature Communications}\ }\textbf {\bibinfo {volume} {4}},\
  \bibinfo {pages} {1951} (\bibinfo {year} {2013})}\BibitemShut {NoStop}%
\bibitem [{\citenamefont {Muravev}\ \emph
  {et~al.}(2016{\natexlab{a}})\citenamefont {Muravev}, \citenamefont
  {Fortunatov}, \citenamefont {Dremin},\ and\ \citenamefont
  {Kukushkin}}]{muravev2016_interfermometer}%
  \BibitemOpen
  \bibfield  {author} {\bibinfo {author} {\bibfnamefont {V.~M.}\ \bibnamefont
  {Muravev}}, \bibinfo {author} {\bibfnamefont {A.~A.}\ \bibnamefont
  {Fortunatov}}, \bibinfo {author} {\bibfnamefont {A.}~\bibnamefont {Dremin}},\
  and\ \bibinfo {author} {\bibfnamefont {I.~V.}\ \bibnamefont {Kukushkin}},\
  }\bibfield  {title} {\bibinfo {title} {Plasmonic interferometer for
  spectroscopy of microwave radiation},\ }\href
  {https://link.springer.com/article/10.1134/S0021364016060084} {\bibfield
  {journal} {\bibinfo  {journal} {JETP Letters}\ }\textbf {\bibinfo {volume}
  {103}},\ \bibinfo {pages} {380} (\bibinfo {year}
  {2016}{\natexlab{a}})}\BibitemShut {NoStop}%
\bibitem [{\citenamefont {Chudow}\ \emph {et~al.}(2016)\citenamefont {Chudow},
  \citenamefont {Santavicca},\ and\ \citenamefont
  {Prober}}]{Chudow_Plasmons_CNTs}%
  \BibitemOpen
  \bibfield  {author} {\bibinfo {author} {\bibfnamefont {J.~D.}\ \bibnamefont
  {Chudow}}, \bibinfo {author} {\bibfnamefont {D.~F.}\ \bibnamefont
  {Santavicca}},\ and\ \bibinfo {author} {\bibfnamefont {D.~E.}\ \bibnamefont
  {Prober}},\ }\bibfield  {title} {\bibinfo {title} {Terahertz spectroscopy of
  individual single-walled carbon nanotubes as a probe of luttinger liquid
  physics},\ }\href {https://doi.org/10.1021/acs.nanolett.6b01485} {\bibfield
  {journal} {\bibinfo  {journal} {Nano Letters}\ }\textbf {\bibinfo {volume}
  {16}},\ \bibinfo {pages} {4909} (\bibinfo {year} {2016})}\BibitemShut
  {NoStop}%
\bibitem [{\citenamefont {Bandurin}\ \emph {et~al.}(2018)\citenamefont
  {Bandurin}, \citenamefont {Svintsov}, \citenamefont {Gayduchenko},
  \citenamefont {Xu}, \citenamefont {Principi}, \citenamefont {Moskotin},
  \citenamefont {Tretyakov}, \citenamefont {Yagodkin}, \citenamefont {Zhukov},
  \citenamefont {Taniguchi}, \citenamefont {Watanabe}, \citenamefont
  {Grigorieva}, \citenamefont {Polini}, \citenamefont {Goltsman}, \citenamefont
  {Geim},\ and\ \citenamefont {Fedorov}}]{Bandurin_resonant}%
  \BibitemOpen
  \bibfield  {author} {\bibinfo {author} {\bibfnamefont {D.~A.}\ \bibnamefont
  {Bandurin}}, \bibinfo {author} {\bibfnamefont {D.}~\bibnamefont {Svintsov}},
  \bibinfo {author} {\bibfnamefont {I.}~\bibnamefont {Gayduchenko}}, \bibinfo
  {author} {\bibfnamefont {S.~G.}\ \bibnamefont {Xu}}, \bibinfo {author}
  {\bibfnamefont {A.}~\bibnamefont {Principi}}, \bibinfo {author}
  {\bibfnamefont {M.}~\bibnamefont {Moskotin}}, \bibinfo {author}
  {\bibfnamefont {I.}~\bibnamefont {Tretyakov}}, \bibinfo {author}
  {\bibfnamefont {D.}~\bibnamefont {Yagodkin}}, \bibinfo {author}
  {\bibfnamefont {S.}~\bibnamefont {Zhukov}}, \bibinfo {author} {\bibfnamefont
  {T.}~\bibnamefont {Taniguchi}}, \bibinfo {author} {\bibfnamefont
  {K.}~\bibnamefont {Watanabe}}, \bibinfo {author} {\bibfnamefont {I.~V.}\
  \bibnamefont {Grigorieva}}, \bibinfo {author} {\bibfnamefont
  {M.}~\bibnamefont {Polini}}, \bibinfo {author} {\bibfnamefont {G.~N.}\
  \bibnamefont {Goltsman}}, \bibinfo {author} {\bibfnamefont {A.~K.}\
  \bibnamefont {Geim}},\ and\ \bibinfo {author} {\bibfnamefont
  {G.}~\bibnamefont {Fedorov}},\ }\bibfield  {title} {\bibinfo {title}
  {{Resonant terahertz detection using graphene plasmons}},\ }\href
  {https://doi.org/10.1038/s41467-018-07848-w} {\bibfield  {journal} {\bibinfo
  {journal} {Nature Communications}\ }\textbf {\bibinfo {volume} {9}},\
  \bibinfo {pages} {5392} (\bibinfo {year} {2018})}\BibitemShut {NoStop}%
\bibitem [{\citenamefont {Castilla}\ \emph {et~al.}(2019)\citenamefont
  {Castilla}, \citenamefont {Terr{\'{e}}s}, \citenamefont {Autore},
  \citenamefont {Viti}, \citenamefont {Li}, \citenamefont {Nikitin},
  \citenamefont {Vangelidis}, \citenamefont {Watanabe}, \citenamefont
  {Taniguchi}, \citenamefont {Lidorikis}, \citenamefont {Vitiello},
  \citenamefont {Hillenbrand}, \citenamefont {Tielrooij},\ and\ \citenamefont
  {Koppens}}]{Castilla2019}%
  \BibitemOpen
  \bibfield  {author} {\bibinfo {author} {\bibfnamefont {S.}~\bibnamefont
  {Castilla}}, \bibinfo {author} {\bibfnamefont {B.}~\bibnamefont
  {Terr{\'{e}}s}}, \bibinfo {author} {\bibfnamefont {M.}~\bibnamefont
  {Autore}}, \bibinfo {author} {\bibfnamefont {L.}~\bibnamefont {Viti}},
  \bibinfo {author} {\bibfnamefont {J.}~\bibnamefont {Li}}, \bibinfo {author}
  {\bibfnamefont {A.~Y.}\ \bibnamefont {Nikitin}}, \bibinfo {author}
  {\bibfnamefont {I.}~\bibnamefont {Vangelidis}}, \bibinfo {author}
  {\bibfnamefont {K.}~\bibnamefont {Watanabe}}, \bibinfo {author}
  {\bibfnamefont {T.}~\bibnamefont {Taniguchi}}, \bibinfo {author}
  {\bibfnamefont {E.}~\bibnamefont {Lidorikis}}, \bibinfo {author}
  {\bibfnamefont {M.~S.}\ \bibnamefont {Vitiello}}, \bibinfo {author}
  {\bibfnamefont {R.}~\bibnamefont {Hillenbrand}}, \bibinfo {author}
  {\bibfnamefont {K.~J.}\ \bibnamefont {Tielrooij}},\ and\ \bibinfo {author}
  {\bibfnamefont {F.~H.}\ \bibnamefont {Koppens}},\ }\bibfield  {title}
  {\bibinfo {title} {{Fast and Sensitive Terahertz Detection Using an
  Antenna-Integrated Graphene pn Junction}},\ }\href
  {https://doi.org/10.1021/acs.nanolett.8b04171} {\bibfield  {journal}
  {\bibinfo  {journal} {Nano Letters}\ }\textbf {\bibinfo {volume} {19}},\
  \bibinfo {pages} {2765} (\bibinfo {year} {2019})}\BibitemShut {NoStop}%
\bibitem [{\citenamefont {Viti}\ \emph {et~al.}(2020)\citenamefont {Viti},
  \citenamefont {Purdie}, \citenamefont {Lombardo}, \citenamefont {Ferrari},\
  and\ \citenamefont {Vitiello}}]{Viti_hBN_encapsulated_Low_noise}%
  \BibitemOpen
  \bibfield  {author} {\bibinfo {author} {\bibfnamefont {L.}~\bibnamefont
  {Viti}}, \bibinfo {author} {\bibfnamefont {D.~G.}\ \bibnamefont {Purdie}},
  \bibinfo {author} {\bibfnamefont {A.}~\bibnamefont {Lombardo}}, \bibinfo
  {author} {\bibfnamefont {A.~C.}\ \bibnamefont {Ferrari}},\ and\ \bibinfo
  {author} {\bibfnamefont {M.~S.}\ \bibnamefont {Vitiello}},\ }\bibfield
  {title} {\bibinfo {title} {Hbn-encapsulated, graphene-based, room-temperature
  terahertz receivers, with high speed and low noise},\ }\href
  {https://doi.org/10.1021/acs.nanolett.9b05207} {\bibfield  {journal}
  {\bibinfo  {journal} {Nano Letters}\ }\textbf {\bibinfo {volume} {20}},\
  \bibinfo {pages} {3169} (\bibinfo {year} {2020})}\BibitemShut {NoStop}%
\bibitem [{\citenamefont {Gayduchenko}\ \emph {et~al.}(2021)\citenamefont
  {Gayduchenko}, \citenamefont {Xu}, \citenamefont {Alymov}, \citenamefont
  {Moskotin}, \citenamefont {Tretyakov}, \citenamefont {Taniguchi},
  \citenamefont {Watanabe}, \citenamefont {Goltsman}, \citenamefont {Geim},
  \citenamefont {Fedorov}, \citenamefont {Svintsov},\ and\ \citenamefont
  {Bandurin}}]{Gayduchenko2021}%
  \BibitemOpen
  \bibfield  {author} {\bibinfo {author} {\bibfnamefont {I.}~\bibnamefont
  {Gayduchenko}}, \bibinfo {author} {\bibfnamefont {S.~G.}\ \bibnamefont {Xu}},
  \bibinfo {author} {\bibfnamefont {G.}~\bibnamefont {Alymov}}, \bibinfo
  {author} {\bibfnamefont {M.}~\bibnamefont {Moskotin}}, \bibinfo {author}
  {\bibfnamefont {I.}~\bibnamefont {Tretyakov}}, \bibinfo {author}
  {\bibfnamefont {T.}~\bibnamefont {Taniguchi}}, \bibinfo {author}
  {\bibfnamefont {K.}~\bibnamefont {Watanabe}}, \bibinfo {author}
  {\bibfnamefont {G.}~\bibnamefont {Goltsman}}, \bibinfo {author}
  {\bibfnamefont {A.~K.}\ \bibnamefont {Geim}}, \bibinfo {author}
  {\bibfnamefont {G.}~\bibnamefont {Fedorov}}, \bibinfo {author} {\bibfnamefont
  {D.}~\bibnamefont {Svintsov}},\ and\ \bibinfo {author} {\bibfnamefont
  {D.~A.}\ \bibnamefont {Bandurin}},\ }\bibfield  {title} {\bibinfo {title}
  {{Tunnel field-effect transistors for sensitive terahertz detection}},\
  }\href {https://doi.org/10.1038/s41467-020-20721-z} {\bibfield  {journal}
  {\bibinfo  {journal} {Nature Communications}\ }\textbf {\bibinfo {volume}
  {12}},\ \bibinfo {pages} {543} (\bibinfo {year} {2021})}\BibitemShut
  {NoStop}%
\bibitem [{\citenamefont {Bauer}\ \emph {et~al.}(2019)\citenamefont {Bauer},
  \citenamefont {Ramer}, \citenamefont {Chevtchenko}, \citenamefont {Osipov},
  \citenamefont {Cibiraite}, \citenamefont {Pralgauskaite}, \citenamefont
  {Ikamas}, \citenamefont {Lisauskas}, \citenamefont {Heinrich}, \citenamefont
  {Krozer},\ and\ \citenamefont {Roskos}}]{Bauer2019}%
  \BibitemOpen
  \bibfield  {author} {\bibinfo {author} {\bibfnamefont {M.}~\bibnamefont
  {Bauer}}, \bibinfo {author} {\bibfnamefont {A.}~\bibnamefont {Ramer}},
  \bibinfo {author} {\bibfnamefont {S.~A.}\ \bibnamefont {Chevtchenko}},
  \bibinfo {author} {\bibfnamefont {K.~Y.}\ \bibnamefont {Osipov}}, \bibinfo
  {author} {\bibfnamefont {D.}~\bibnamefont {Cibiraite}}, \bibinfo {author}
  {\bibfnamefont {S.}~\bibnamefont {Pralgauskaite}}, \bibinfo {author}
  {\bibfnamefont {K.}~\bibnamefont {Ikamas}}, \bibinfo {author} {\bibfnamefont
  {A.}~\bibnamefont {Lisauskas}}, \bibinfo {author} {\bibfnamefont
  {W.}~\bibnamefont {Heinrich}}, \bibinfo {author} {\bibfnamefont
  {V.}~\bibnamefont {Krozer}},\ and\ \bibinfo {author} {\bibfnamefont {H.~G.}\
  \bibnamefont {Roskos}},\ }\bibfield  {title} {\bibinfo {title} {{A
  High-Sensitivity AlGaN/GaN HEMT Terahertz Detector With Integrated Broadband
  Bow-Tie Antenna}},\ }\href {https://doi.org/10.1109/TTHZ.2019.2917782}
  {\bibfield  {journal} {\bibinfo  {journal} {IEEE Transactions on Terahertz
  Science and Technology}\ }\textbf {\bibinfo {volume} {9}},\ \bibinfo {pages}
  {430} (\bibinfo {year} {2019})}\BibitemShut {NoStop}%
\bibitem [{\citenamefont {Tretyakov}(2014)}]{tretyakov2014maximizing}%
  \BibitemOpen
  \bibfield  {author} {\bibinfo {author} {\bibfnamefont {S.}~\bibnamefont
  {Tretyakov}},\ }\bibfield  {title} {\bibinfo {title} {Maximizing absorption
  and scattering by dipole particles},\ }\href
  {https://doi.org/10.1007/s11468-014-9699-y} {\bibfield  {journal} {\bibinfo
  {journal} {Plasmonics}\ }\textbf {\bibinfo {volume} {9}},\ \bibinfo {pages}
  {935} (\bibinfo {year} {2014})}\BibitemShut {NoStop}%
\bibitem [{\citenamefont {Balanis}(2015)}]{balanis2015antenna}%
  \BibitemOpen
  \bibfield  {author} {\bibinfo {author} {\bibfnamefont {C.~A.}\ \bibnamefont
  {Balanis}},\ }\href@noop {} {\emph {\bibinfo {title} {Antenna theory:
  analysis and design}}}\ (\bibinfo  {publisher} {John wiley \& sons},\
  \bibinfo {year} {2015})\BibitemShut {NoStop}%
\bibitem [{\citenamefont {Yesilkoy}\ \emph {et~al.}(2019)\citenamefont
  {Yesilkoy}, \citenamefont {Arvelo}, \citenamefont {Jahani}, \citenamefont
  {Liu}, \citenamefont {Tittl}, \citenamefont {Cevher}, \citenamefont
  {Kivshar},\ and\ \citenamefont {Altug}}]{Kivshar_Hyperspectral}%
  \BibitemOpen
  \bibfield  {author} {\bibinfo {author} {\bibfnamefont {F.}~\bibnamefont
  {Yesilkoy}}, \bibinfo {author} {\bibfnamefont {E.~R.}\ \bibnamefont
  {Arvelo}}, \bibinfo {author} {\bibfnamefont {Y.}~\bibnamefont {Jahani}},
  \bibinfo {author} {\bibfnamefont {M.}~\bibnamefont {Liu}}, \bibinfo {author}
  {\bibfnamefont {A.}~\bibnamefont {Tittl}}, \bibinfo {author} {\bibfnamefont
  {V.}~\bibnamefont {Cevher}}, \bibinfo {author} {\bibfnamefont
  {Y.}~\bibnamefont {Kivshar}},\ and\ \bibinfo {author} {\bibfnamefont
  {H.}~\bibnamefont {Altug}},\ }\bibfield  {title} {\bibinfo {title}
  {{Ultrasensitive hyperspectral imaging and biodetection enabled by dielectric
  metasurfaces}},\ }\href {https://doi.org/10.1038/s41566-019-0394-6}
  {\bibfield  {journal} {\bibinfo  {journal} {Nature Photonics}\ }\textbf
  {\bibinfo {volume} {13}},\ \bibinfo {pages} {390} (\bibinfo {year}
  {2019})}\BibitemShut {NoStop}%
\bibitem [{\citenamefont {Nagatsuma}\ \emph {et~al.}(2016)\citenamefont
  {Nagatsuma}, \citenamefont {Ducournau},\ and\ \citenamefont
  {Renaud}}]{Nagatsuma2016}%
  \BibitemOpen
  \bibfield  {author} {\bibinfo {author} {\bibfnamefont {T.}~\bibnamefont
  {Nagatsuma}}, \bibinfo {author} {\bibfnamefont {G.}~\bibnamefont
  {Ducournau}},\ and\ \bibinfo {author} {\bibfnamefont {C.~C.}\ \bibnamefont
  {Renaud}},\ }\bibfield  {title} {\bibinfo {title} {{Advances in terahertz
  communications accelerated by photonics}},\ }\href
  {https://doi.org/10.1038/nphoton.2016.65} {\bibfield  {journal} {\bibinfo
  {journal} {Nature Photonics}\ }\textbf {\bibinfo {volume} {10}},\ \bibinfo
  {pages} {371} (\bibinfo {year} {2016})}\BibitemShut {NoStop}%
\bibitem [{\citenamefont {Sharma}\ \emph {et~al.}(2015)\citenamefont {Sharma},
  \citenamefont {Singh}, \citenamefont {Bougher},\ and\ \citenamefont
  {Cola}}]{Sharma_rectenna}%
  \BibitemOpen
  \bibfield  {author} {\bibinfo {author} {\bibfnamefont {A.}~\bibnamefont
  {Sharma}}, \bibinfo {author} {\bibfnamefont {V.}~\bibnamefont {Singh}},
  \bibinfo {author} {\bibfnamefont {T.~L.}\ \bibnamefont {Bougher}},\ and\
  \bibinfo {author} {\bibfnamefont {B.~A.}\ \bibnamefont {Cola}},\ }\bibfield
  {title} {\bibinfo {title} {{A carbon nanotube optical rectenna}},\ }\href
  {https://doi.org/10.1038/nnano.2015.220} {\bibfield  {journal} {\bibinfo
  {journal} {Nature Nanotechnology}\ }\textbf {\bibinfo {volume} {10}},\
  \bibinfo {pages} {1027} (\bibinfo {year} {2015})}\BibitemShut {NoStop}%
\bibitem [{\citenamefont {Spinelli}\ \emph {et~al.}(2011)\citenamefont
  {Spinelli}, \citenamefont {Hebbink}, \citenamefont {de~Waele}, \citenamefont
  {Black}, \citenamefont {Lenzmann},\ and\ \citenamefont
  {Polman}}]{Spinelli_ImpedanceMatching_2011}%
  \BibitemOpen
  \bibfield  {author} {\bibinfo {author} {\bibfnamefont {P.}~\bibnamefont
  {Spinelli}}, \bibinfo {author} {\bibfnamefont {M.}~\bibnamefont {Hebbink}},
  \bibinfo {author} {\bibfnamefont {R.}~\bibnamefont {de~Waele}}, \bibinfo
  {author} {\bibfnamefont {L.}~\bibnamefont {Black}}, \bibinfo {author}
  {\bibfnamefont {F.}~\bibnamefont {Lenzmann}},\ and\ \bibinfo {author}
  {\bibfnamefont {A.}~\bibnamefont {Polman}},\ }\bibfield  {title} {\bibinfo
  {title} {{Optical Impedance Matching Using Coupled Plasmonic Nanoparticle
  Arrays}},\ }\href {https://doi.org/10.1021/nl200321u} {\bibfield  {journal}
  {\bibinfo  {journal} {Nano Letters}\ }\textbf {\bibinfo {volume} {11}},\
  \bibinfo {pages} {1760} (\bibinfo {year} {2011})}\BibitemShut {NoStop}%
\bibitem [{\citenamefont {Ginzburg}\ and\ \citenamefont
  {Orenstein}(2007)}]{Ginzburg_impedance_matching_2007}%
  \BibitemOpen
  \bibfield  {author} {\bibinfo {author} {\bibfnamefont {P.}~\bibnamefont
  {Ginzburg}}\ and\ \bibinfo {author} {\bibfnamefont {M.}~\bibnamefont
  {Orenstein}},\ }\bibfield  {title} {\bibinfo {title} {{Plasmonic transmission
  lines: from micro to nano scale with $\lambda$/4 impedance matching}},\
  }\href {https://doi.org/10.1364/OE.15.006762} {\bibfield  {journal} {\bibinfo
   {journal} {Optics Express}\ }\textbf {\bibinfo {volume} {15}},\ \bibinfo
  {pages} {6762} (\bibinfo {year} {2007})}\BibitemShut {NoStop}%
\bibitem [{\citenamefont {Dias}\ and\ \citenamefont {{Garc{\'{i}}a De
  Abajo}}(2019)}]{Abajo_Limits_to_Coupling}%
  \BibitemOpen
  \bibfield  {author} {\bibinfo {author} {\bibfnamefont {E.~J.}\ \bibnamefont
  {Dias}}\ and\ \bibinfo {author} {\bibfnamefont {F.~J.}\ \bibnamefont
  {{Garc{\'{i}}a De Abajo}}},\ }\bibfield  {title} {\bibinfo {title}
  {{Fundamental Limits to the Coupling between Light and 2D Polaritons by Small
  Scatterers}},\ }\href {https://doi.org/10.1021/acsnano.8b09283} {\bibfield
  {journal} {\bibinfo  {journal} {ACS Nano}\ }\textbf {\bibinfo {volume}
  {13}},\ \bibinfo {pages} {5184} (\bibinfo {year} {2019})}\BibitemShut
  {NoStop}%
\bibitem [{\citenamefont {Zagorodnev}\ \emph {et~al.}(2021)\citenamefont
  {Zagorodnev}, \citenamefont {Rodionov},\ and\ \citenamefont
  {Zabolotnykh}}]{Zagorodnev-Effect-of-retardation}%
  \BibitemOpen
  \bibfield  {author} {\bibinfo {author} {\bibfnamefont {I.~V.}\ \bibnamefont
  {Zagorodnev}}, \bibinfo {author} {\bibfnamefont {D.~A.}\ \bibnamefont
  {Rodionov}},\ and\ \bibinfo {author} {\bibfnamefont {A.~A.}\ \bibnamefont
  {Zabolotnykh}},\ }\bibfield  {title} {\bibinfo {title} {Effect of retardation
  on the frequency and linewidth of plasma resonances in a two-dimensional disk
  of electron gas},\ }\href {https://doi.org/10.1103/PhysRevB.103.195431}
  {\bibfield  {journal} {\bibinfo  {journal} {Phys. Rev. B}\ }\textbf {\bibinfo
  {volume} {103}},\ \bibinfo {pages} {195431} (\bibinfo {year}
  {2021})}\BibitemShut {NoStop}%
\bibitem [{\citenamefont {Kuang}\ \emph {et~al.}(2020)\citenamefont {Kuang},
  \citenamefont {Zhang},\ and\ \citenamefont
  {Miller}}]{Kuang_MaxOpticalResponse}%
  \BibitemOpen
  \bibfield  {author} {\bibinfo {author} {\bibfnamefont {Z.}~\bibnamefont
  {Kuang}}, \bibinfo {author} {\bibfnamefont {L.}~\bibnamefont {Zhang}},\ and\
  \bibinfo {author} {\bibfnamefont {O.~D.}\ \bibnamefont {Miller}},\ }\bibfield
   {title} {\bibinfo {title} {Maximal single-frequency electromagnetic
  response},\ }\href {https://doi.org/10.1364/OPTICA.398715} {\bibfield
  {journal} {\bibinfo  {journal} {Optica}\ }\textbf {\bibinfo {volume} {7}},\
  \bibinfo {pages} {1746} (\bibinfo {year} {2020})}\BibitemShut {NoStop}%
\bibitem [{\citenamefont {Miller}\ \emph {et~al.}(2017)\citenamefont {Miller},
  \citenamefont {Ilic}, \citenamefont {Christensen}, \citenamefont {Reid},
  \citenamefont {Atwater}, \citenamefont {Joannopoulos}, \citenamefont
  {Soljačić},\ and\ \citenamefont {Johnson}}]{Miller_MaxOpticalResponse2dm}%
  \BibitemOpen
  \bibfield  {author} {\bibinfo {author} {\bibfnamefont {O.~D.}\ \bibnamefont
  {Miller}}, \bibinfo {author} {\bibfnamefont {O.}~\bibnamefont {Ilic}},
  \bibinfo {author} {\bibfnamefont {T.}~\bibnamefont {Christensen}}, \bibinfo
  {author} {\bibfnamefont {M.~T.~H.}\ \bibnamefont {Reid}}, \bibinfo {author}
  {\bibfnamefont {H.~A.}\ \bibnamefont {Atwater}}, \bibinfo {author}
  {\bibfnamefont {J.~D.}\ \bibnamefont {Joannopoulos}}, \bibinfo {author}
  {\bibfnamefont {M.}~\bibnamefont {Soljačić}},\ and\ \bibinfo {author}
  {\bibfnamefont {S.~G.}\ \bibnamefont {Johnson}},\ }\bibfield  {title}
  {\bibinfo {title} {Limits to the optical response of graphene and
  two-dimensional materials},\ }\href
  {https://doi.org/10.1021/acs.nanolett.7b02007} {\bibfield  {journal}
  {\bibinfo  {journal} {Nano Letters}\ }\textbf {\bibinfo {volume} {17}},\
  \bibinfo {pages} {5408} (\bibinfo {year} {2017})}\BibitemShut {NoStop}%
\bibitem [{Note1()}]{Note1}%
  \BibitemOpen
  \bibinfo {note} {For GaAs quantum well with $m^*=0.067m_0$ this inductance
  corresponds to sheet density $n_s = 1.2 \times 10^{13}$ cm$^{-2}$, while
  $\tau = 0.1$ ps corresponds to mobility $\mu \approx 2600$ cm$^2$/V
  s.}\BibitemShut {Stop}%
\bibitem [{\citenamefont {Muravev}\ \emph
  {et~al.}(2016{\natexlab{b}})\citenamefont {Muravev}, \citenamefont {Andreev},
  \citenamefont {Gubarev}, \citenamefont {Belyanin},\ and\ \citenamefont
  {Kukushkin}}]{Muravev_FineStructureCR}%
  \BibitemOpen
  \bibfield  {author} {\bibinfo {author} {\bibfnamefont {V.~M.}\ \bibnamefont
  {Muravev}}, \bibinfo {author} {\bibfnamefont {I.~V.}\ \bibnamefont
  {Andreev}}, \bibinfo {author} {\bibfnamefont {S.~I.}\ \bibnamefont
  {Gubarev}}, \bibinfo {author} {\bibfnamefont {V.~N.}\ \bibnamefont
  {Belyanin}},\ and\ \bibinfo {author} {\bibfnamefont {I.~V.}\ \bibnamefont
  {Kukushkin}},\ }\bibfield  {title} {\bibinfo {title} {Fine structure of
  cyclotron resonance in a two-dimensional electron system},\ }\href
  {https://doi.org/10.1103/PhysRevB.93.041110} {\bibfield  {journal} {\bibinfo
  {journal} {Phys. Rev. B}\ }\textbf {\bibinfo {volume} {93}},\ \bibinfo
  {pages} {041110} (\bibinfo {year} {2016}{\natexlab{b}})}\BibitemShut
  {NoStop}%
\bibitem [{\citenamefont {Muravev}\ \emph {et~al.}(2015)\citenamefont
  {Muravev}, \citenamefont {Gusikhin}, \citenamefont {Andreev},\ and\
  \citenamefont {Kukushkin}}]{Muravev_PRL_Relativistic}%
  \BibitemOpen
  \bibfield  {author} {\bibinfo {author} {\bibfnamefont {V.~M.}\ \bibnamefont
  {Muravev}}, \bibinfo {author} {\bibfnamefont {P.~A.}\ \bibnamefont
  {Gusikhin}}, \bibinfo {author} {\bibfnamefont {I.~V.}\ \bibnamefont
  {Andreev}},\ and\ \bibinfo {author} {\bibfnamefont {I.~V.}\ \bibnamefont
  {Kukushkin}},\ }\bibfield  {title} {\bibinfo {title} {Novel relativistic
  plasma excitations in a gated two-dimensional electron system},\ }\href
  {https://doi.org/10.1103/PhysRevLett.114.106805} {\bibfield  {journal}
  {\bibinfo  {journal} {Phys. Rev. Lett.}\ }\textbf {\bibinfo {volume} {114}},\
  \bibinfo {pages} {106805} (\bibinfo {year} {2015})}\BibitemShut {NoStop}%
\bibitem [{\citenamefont {Muravev}\ \emph
  {et~al.}(2020{\natexlab{a}})\citenamefont {Muravev}, \citenamefont
  {Gusikhin}, \citenamefont {Zarezin}, \citenamefont {Zabolotnykh},
  \citenamefont {Volkov},\ and\ \citenamefont
  {Kukushkin}}]{Muravev_PhysicalOrigin}%
  \BibitemOpen
  \bibfield  {author} {\bibinfo {author} {\bibfnamefont {V.~M.}\ \bibnamefont
  {Muravev}}, \bibinfo {author} {\bibfnamefont {P.~A.}\ \bibnamefont
  {Gusikhin}}, \bibinfo {author} {\bibfnamefont {A.~M.}\ \bibnamefont
  {Zarezin}}, \bibinfo {author} {\bibfnamefont {A.~A.}\ \bibnamefont
  {Zabolotnykh}}, \bibinfo {author} {\bibfnamefont {V.~A.}\ \bibnamefont
  {Volkov}},\ and\ \bibinfo {author} {\bibfnamefont {I.~V.}\ \bibnamefont
  {Kukushkin}},\ }\bibfield  {title} {\bibinfo {title} {{Physical origin of
  relativistic plasmons in a two-dimensional electron system}},\ }\href
  {https://doi.org/10.1103/physrevb.102.081301} {\bibfield  {journal} {\bibinfo
   {journal} {Physical Review B}\ }\textbf {\bibinfo {volume} {102}},\ \bibinfo
  {pages} {81301} (\bibinfo {year} {2020}{\natexlab{a}})}\BibitemShut {NoStop}%
\bibitem [{\citenamefont {Muravev}\ \emph
  {et~al.}(2020{\natexlab{b}})\citenamefont {Muravev}, \citenamefont {Semenov},
  \citenamefont {Andreev}, \citenamefont {Gusikhin},\ and\ \citenamefont
  {Kukushkin}}]{Muravev_LCmode}%
  \BibitemOpen
  \bibfield  {author} {\bibinfo {author} {\bibfnamefont {V.~M.}\ \bibnamefont
  {Muravev}}, \bibinfo {author} {\bibfnamefont {N.~D.}\ \bibnamefont
  {Semenov}}, \bibinfo {author} {\bibfnamefont {I.~V.}\ \bibnamefont
  {Andreev}}, \bibinfo {author} {\bibfnamefont {P.~A.}\ \bibnamefont
  {Gusikhin}},\ and\ \bibinfo {author} {\bibfnamefont {I.~V.}\ \bibnamefont
  {Kukushkin}},\ }\bibfield  {title} {\bibinfo {title} {{A tunable plasmonic
  resonator using kinetic 2D inductance and patch capacitance}},\ }\href
  {https://doi.org/10.1063/5.0026034} {\bibfield  {journal} {\bibinfo
  {journal} {Applied Physics Letters}\ }\textbf {\bibinfo {volume} {117}},\
  \bibinfo {pages} {151103} (\bibinfo {year} {2020}{\natexlab{b}})}\BibitemShut
  {NoStop}%
\bibitem [{\citenamefont {Echtermeyer}\ \emph {et~al.}(2011)\citenamefont
  {Echtermeyer}, \citenamefont {Britnell}, \citenamefont {Jasnos},
  \citenamefont {Lombardo}, \citenamefont {Gorbachev}, \citenamefont
  {Grigorenko}, \citenamefont {Geim}, \citenamefont {Ferrari},\ and\
  \citenamefont {Novoselov}}]{Echtermeyer2011}%
  \BibitemOpen
  \bibfield  {author} {\bibinfo {author} {\bibfnamefont {T.}~\bibnamefont
  {Echtermeyer}}, \bibinfo {author} {\bibfnamefont {L.}~\bibnamefont
  {Britnell}}, \bibinfo {author} {\bibfnamefont {P.}~\bibnamefont {Jasnos}},
  \bibinfo {author} {\bibfnamefont {A.}~\bibnamefont {Lombardo}}, \bibinfo
  {author} {\bibfnamefont {R.}~\bibnamefont {Gorbachev}}, \bibinfo {author}
  {\bibfnamefont {A.}~\bibnamefont {Grigorenko}}, \bibinfo {author}
  {\bibfnamefont {A.}~\bibnamefont {Geim}}, \bibinfo {author} {\bibfnamefont
  {A.}~\bibnamefont {Ferrari}},\ and\ \bibinfo {author} {\bibfnamefont
  {K.}~\bibnamefont {Novoselov}},\ }\bibfield  {title} {\bibinfo {title}
  {{Strong plasmonic enhancement of photovoltage in graphene}},\ }\href
  {https://doi.org/10.1038/ncomms1464} {\bibfield  {journal} {\bibinfo
  {journal} {Nature Communications}\ }\textbf {\bibinfo {volume} {2}},\
  \bibinfo {pages} {458} (\bibinfo {year} {2011})},\ \Eprint
  {https://arxiv.org/abs/1107.4176} {1107.4176} \BibitemShut {NoStop}%
\bibitem [{\citenamefont {Wang}\ \emph {et~al.}(2020)\citenamefont {Wang},
  \citenamefont {Allcca}, \citenamefont {Chung}, \citenamefont {Kildishev},
  \citenamefont {Chen}, \citenamefont {Boltasseva},\ and\ \citenamefont
  {Shalaev}}]{Shalaev_2020}%
  \BibitemOpen
  \bibfield  {author} {\bibinfo {author} {\bibfnamefont {D.}~\bibnamefont
  {Wang}}, \bibinfo {author} {\bibfnamefont {A.~E.~L.}\ \bibnamefont {Allcca}},
  \bibinfo {author} {\bibfnamefont {T.-F.}\ \bibnamefont {Chung}}, \bibinfo
  {author} {\bibfnamefont {A.~V.}\ \bibnamefont {Kildishev}}, \bibinfo {author}
  {\bibfnamefont {Y.~P.}\ \bibnamefont {Chen}}, \bibinfo {author}
  {\bibfnamefont {A.}~\bibnamefont {Boltasseva}},\ and\ \bibinfo {author}
  {\bibfnamefont {V.~M.}\ \bibnamefont {Shalaev}},\ }\bibfield  {title}
  {\bibinfo {title} {{Enhancing the graphene photocurrent using surface
  plasmons and a p-n junction}},\ }\href
  {https://doi.org/10.1038/s41377-020-00344-1} {\bibfield  {journal} {\bibinfo
  {journal} {Light: Science \& Applications}\ }\textbf {\bibinfo {volume}
  {9}},\ \bibinfo {pages} {126} (\bibinfo {year} {2020})}\BibitemShut {NoStop}%
\end{thebibliography}%

\end{document}